\documentclass[pra,aps,showpacs, twocolumn, floatfix,longbibliography]{revtex4-1} %%
\usepackage {graphicx} %%
\usepackage {amssymb} %%
\usepackage {amsmath}
\usepackage [mathcal]{euscript}
\usepackage {upgreek}
\usepackage {color}
\usepackage{fdsymbol}
\newcommand{\reffig}[1]{Fig.~\ref{#1}}
\newcommand{\refeq}[1]{Eq.~(\ref{#1})}
\newcommand{\refeqs}[2]{Eqs.~(\ref{#1})-(\ref{#2})}

\DeclareMathOperator{\atanh}{atanh}
\DeclareMathOperator{\sign}{sign}

\newcommand*{\lar}{{\mapsfrom}}
\newcommand*{\rar}{{\mapsto}}
\newcommand{\bra}[1]{\ensuremath{\left\langle#1\right|}}
\newcommand{\ket}[1]{\ensuremath{\left|#1\right\rangle}}

\usepackage[markup=default]{changes}
\setremarkmarkup{\textbf{[[#1:\textcolor{Changes@Color#1}{#2}]]}}

\begin{document}

%%%%     ver 4 title
\title{Weak stochastic ratchets and dynamic localization in measurement-induced quantum trajectories}

% valid only in PR
\author{I. Babushkin}
\affiliation{Institute of Quantum Optics, Leibniz University Hannover,
  Welfengarten 1, 30167 Hannover, Germany} 
\affiliation{Max Born Institute, Max-Born-Strasse 2a, 10117, Berlin, Germany}

\date{\today}

\begin{abstract}
  We consider a qubit governed by a sequence of weak measurements,
  with the measurement strength modified in a time- and
  state-dependent manner.  The resulting trajectory of the qubit in
  the phase space can be weakly controlled without any direct action
  on the qubit (control-free control), even only one fixed observable
  is measured.  Here we show a possibility of a weak form of a
  stochastic ratchet, allowing to create an additional ``force''
  without changing the corresponding effective average potential.
  Furthermore, if the weak measurement strength is significantly
  reduced in a way conditioned to some particular state, a dynamical
  localization near this state takes place. If the measurement
  strength is reduced to zero, a singularity appears, which behaves
  like an artificial basis state.
\end{abstract}
%%%%%%%%%

% \begin{keyword}
% Quantum Weak Zeno Measurements \sep Random Walks \sep Brownian Motion

%
% 03.65.Xp Tunneling, traversal time, quantum Zeno dynamics
%  03.67.-a  Quantum information (see also 42.50.Dv Quantum state engineering and measurements; 42.50.Ex Optical implementations of quantum information processing and transfer in quantum optics)
%
%  03.65.Ta Foundations of quantum mechanics; measurement theory (for
%  optical tests of quantum theory, see 42.50.Xa)
% 05.40.Fb Random walks and Levy flights
% 05.40.Jc	Brownian motion

% \PACS  42.50.Dv \sep 05.40.Fb \sep 03.65.Xp \sep 03.65.Ta \sep 05.40.Jc
% \end{keyword}


\maketitle

%\end{frontmatter}

\section{\label{sec:intro}Introduction}

Dynamics of quantum systems with the states frequently monitored by a
measurement apparatus can be rather controversial and is a subject of
constant interest over the years
\cite{patil15,blok14,murch13,hatridge13,wiseman11,ashhab10,guerlin07,gleyzes07,murch08,gammelmark13,pechen06,gordon13,gammelmark13,
  mackrory10,wiseman10:book,paz-silva12}.  Every measurement in the
measurement sequence may be tuned to cause only a partial collapse of the system's
wave function, with the collapse effect being arbitrary weak (so called
``weak measurements'') \cite{hatridge13,
  gleyzes07,caves87,oreshkov05,varbanov07}. %
In another context, the term ``weak measurement'' was introduced by
Aharonov, Albert, and Vaidman (AAV) \cite{aharonov88} as being
attributed to a measurement of a continuous degree of freedom
(e.~g. an electron position) coupled to a discrete one (spin) via
post-selection.  These two approaches were recently shown to be
equivalent \cite{dressel12,lundeen12}.
As a result of repetitive application of weak measurements --- the
situation which is sometimes referred to as weak Zeno measurements
(WZM) --- a kind of stochastic ``quantum trajectory'' arises
\cite{murch13,belavkin89,gisin84,wiseman96,wiseman10:book} due to
unpredictable character of every particular measurement outcome.
 
Just repeating the weak measurements, without any further action on
the system, allows to control the system state in various ways. For
instance, one can achieve an arbitrary state from any other one by
repeating weak measurements in different bases
\cite{ashhab10,gillett10,mackrory10,wiseman11,karasik11,gordon13,
  blok14}.  Alternatively, by allowing the strength of the weak
measurements (in AAV sense) to depend on the spatial coordinate, one
can create a ``potential wall'' which may reflect a particle
\cite{mackrory10,gordon13}. It should be noted that these control
mechanisms work if the initial state is pare-known.

In contrast, if we repeat a weak measurement of a single qubit in a
fixed basis, the resulting dynamics was up to now believed to be very
trivial. The resulting quantum trajectory just stochastically
approaches one of the two qubit's basis states $\ket{0}$ or $\ket{1}$.
In this article we add a new dimension to this seemingly trivial
dynamics by allowing the measurement strength to be changed in time
and in a state-conditional way.

We observe that the stochastic
equations, describing quantum trajectories in such simple
one-dimensional system are in fact quite similar to the ones describing 
motion of a small Brownian particle in a fluid flow (overdamped
Brownian motion)
\cite{gisin84,haenggi09,reimann02,braun:book04}.
One of the striking phenomena in such  flows as well as in many
other stochastic systems are so called stochastic ratchets. Namely, by
varying the potential acting on the Brownian particle in time and/or
space in a periodic way, it is possible to create an effective
additional force, despite the potential introduces no average force
\cite{haenggi09,reimann02,braun:book04}.  
Such ratchets are encountered in many stochastic
systems of different nature
\cite{wu14,haenggi09,velez08,souza06,villegas05,reimann02,braun:book04}.
Brownian ratchets are deeply connected to so called
Parrondo games, when two or more lossy games are
combined to give a winning one
\cite{parrondo96,allison02,wu14}. Very recently, 
the notion of weak Parrondo games and weak Brownian
ratchets were introduced in \cite{wu14} to describe the situation when
two or more lossy games are played together to give just less lossy (but 
not winning) one.
We remark that, although both Parrondo games and Brownian ratchets
were considered in context of quantum systems
\cite{pawela13,chandrashekar11,bulger08,flitney02,reimann02}, this was
up to now done via some direct action on the system itself.

In contrast,
in the present article
  the situation of control-without-direct-action
will be discussed.
Here we show how the effective potential arising in WZM dynamics can be
modified by changing the strength of the measurement periodically in
time and in a state-conditioned fashion (that is, in dependence on the
current system state).  We demonstrate that the stochastic ratchet
effect, albeit weak, is possible in such situation.  We
also demonstrate a dynamic
localization of the state in the case when the measurement strength
 vanishes at some particular system state. 
 A ``false basis state'' may appear, which ``attracts'' the stochastic
 trajectories in similar way as the true basis states do.

The article is organized as follows: in Sec.~\ref{sec:discr} we
introduce the system under consideration and derive the master
equation governing the probability distribution on the line between
two basis states; In Sec.~\ref{sec:contin} we derive the equation for
the continuous case taking into account conditionally-dependent
measurement strength; In Secs.~\ref{sec:ratchets}-\ref{sec:localiz}
we investigate the effects related to the conditionally modified
measurements strength. Finally, the
conclusions and discussion are presented in Sec.~\ref{sec:concl}.

\section{Discrete dynamics}
\label{sec:discr}

\begin{figure*}[ht]%[htbp!]
  \begin{center}
    \includegraphics[width=0.75\textwidth]{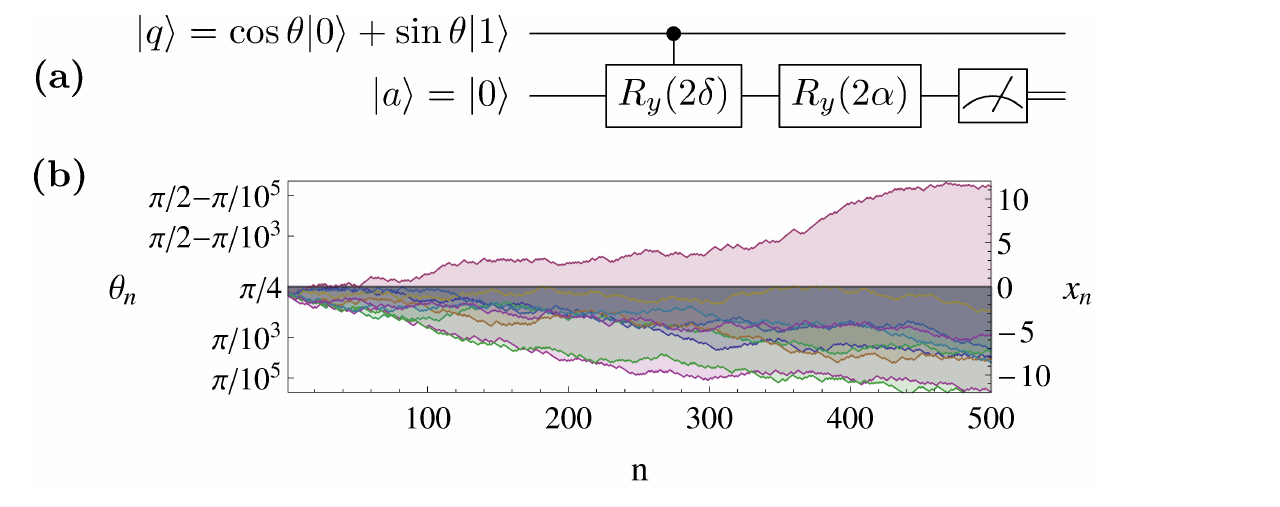}%%
    \caption{ \label{f1}%%
      (a) The model of weak measurements of the qubit $\ket{q}$ using
      an ancilla $\ket{a}$ (initially in the state $\ket{0}$) and its
      rotations $R_y$, followed by the measurement of $\ket{a}$. After
      the measurement, the process is repeated with the same or
      another ancilla in the state $\ket{0}$. In (b), the stochastic
      process induced by repeating application of (a) using $\theta$
      coordinates (left y-axis) and $x$-coordinates given by
      \refeq{eq:1} (right y-axis) vs. measurement number $n$ is shown.
      The $x$-coordinates are obviously better suitable to study the
      asymptotic behavior as $n\to\infty$.  }
  \end{center}
\end{figure*}

\subsection{The setting}

Our model of weak Zeno measurements is
depictured in \reffig{f1}(a) and uses projective measurements of
ancilla qubit $\ket{a}$ to realize the weak ones of $\ket{q}$. The system is prepared in the state $\ket{q}\otimes\ket{0}$
with $\ket{q} = \cos\theta \ket{0} + \sin\theta\ket{1}$ for some
$\theta$. First, we apply a rotation $R_y(2\delta)$ to
the ancilla state; the rotation $R_y$ is conditioned to
$\ket{q}=\ket{1}$ and is defined as: 
\begin{gather}R_y(\delta)\ket{0} \mapsto \cos\delta \ket{0} +
  \sin\delta\ket{1}, \\R_y(\delta)\ket{1} \mapsto \cos\delta \ket{1} -
  \sin\delta\ket{0}.\end{gather}
Afterwards, we  unconditionally apply the rotation $R_y$ by %rotate by
some other angle $\alpha$ and finally we measure the ancilla qubit. If
$\alpha\ne0$ and  $\delta$ is small, the resulting measurement modifies
$\ket{q}$ only slightly. After the measurement, we repeat the whole procedure
using another ancilla in the initial state $\ket{0}$ (or the same
ancilla returned to the state $\ket{0}$). 
%

%%%%%%%%%%%%%
The  state of the whole system   
$\ket{q}\otimes\ket{0}$ [see \reffig{f1}(a)] is  transformed by two
$R_y$ operators described above as:
%% eq WBXVIII:13.02.13
\begin{gather}
\label{eq:0}
     \ket{q}\otimes\ket{0} \mapsto \ket{q_0} \otimes \ket{0} + \ket{q_1}
  \otimes \ket{1}, \\  \ket{q_i} = \sum_j b_{ij} \ket{j}, 
\end{gather}
where $b_{ij}$ are $(i+1,j+1)$th element of the matrix $b$ defined as:
\begin{equation}
  \label{eq:4}
    b =
\begin{pmatrix}
  & \cos\theta\cos\alpha & \sin\theta\cos{(\delta +\alpha)} \\ 
  & \cos\theta\sin\alpha & \sin\theta\sin{(\delta+ \alpha )} \\
\end{pmatrix}. 
\end{equation}

If the measurement of $\ket{a}$ gives 0, the system state is
reduced to $\ket{q} = \ket{q_0}/\sqrt{p_0}$ and 
in the opposite case to $\ket{q} = \ket{q_1}/\sqrt{p_1}$, where
\begin{gather}
  \label{eq:2}
  p_0 = \cos^2\alpha \cos^2\theta + \sin^2\theta\cos^2{(\delta
    +\alpha)},\\
p_1 = \cos^2\theta\sin^2\alpha + \sin^2\theta\sin^2{(\delta+ \alpha
  )}.
  \label{eq:2a}
\end{gather}
The probabilities of the corresponding outcomes are $p_0$ and
$p_1$. 

The above description can be reformulated in the form of a generalized
measurement formalism, with the measurement operators
\begin{eqnarray}
  \label{eq:6}
  \mathcal{B}_0 = b_{11}\ket{0}\bra{0} + b_{12}\ket{1}\bra{1},\\
  \mathcal{B}_1 = b_{21}\ket{0}\bra{0} + b_{22}\ket{1}\bra{1},
  \label{eq:6a}
\end{eqnarray}
so that $\sum_j \mathcal{B}_j^\dag\mathcal{B}_j =1$ and the state
after the measurement with the result $j$ is transformed as:
$\ket{q}\to \mathcal{B}_j\ket{q}/\sqrt{p_j}$.

The resulting process is a (classical) one-dimensional random walk
along the axis $\theta$ as shown in \reffig{f1}(b). It spends most of
time in the vicinity of the limiting states $\ket{q}=\ket{0}$ and
$\ket{q}=\ket{1}$ ($\theta=0$, $\pi/2$).  It is thus useful to
introduce parabolic coordinates \cite{oreshkov05} as:
\begin{equation}
  \label{eq:1}
 x= \atanh{\left\{-\cos(2\theta)\right\}}; \, \theta = \arcsin{\sqrt{\frac{1+\tanh{x}}{2}}}.
\end{equation}
In this coordinate system, $\theta=0$ corresponds to $x=-\infty$ and
$\theta=\pi/2$ corresponds to $x=+\infty$ [(cf. \reffig{f1}(b)].
Using $x$ instead of $\theta$ allows to expand these vicinities into
semi-infinite intervals.
We also note that if we measure $\ket{q}$ directly (instead of
$\ket{a}$), the probability to find $\ket{q}$ in the state $\ket{1}$
will be:
\begin{equation}
\Pi(x) = \sin^2\theta = (1+\tanh x)/2.\label{eq:pi}
\end{equation}

\subsection{Classical random walk interpretation}

Now, for the sake of simplicity, we exclude the situation when
$\ket{q}$ is exactly in one of the basis states $\ket{0}$ or $\ket{1}$
in the beginning of the process.  In this case, the equations above
allow to define the process as a one-dimensional random walk on the
line $x\in (-\infty,+\infty)$ in the following way: assuming that at
the $n$th iteration step the system is in the point $x_n$, at the
$n+1$ step it will be in the point either
$x_{n+1} = x_{n} + \epsilon_0$ or $x_{n+1} = x_{n} + \epsilon_1$
(depending on the measurement outcome), every of two variants
occurring with the probabilities $p_i(x_n)$, $i=0,1$. Few realizations
of this random walks are shown in \reffig{f1}(b).

The probabilities $p_i(x)$ can be rewritten in $x$-coordinates as:
\begin{gather}
  \label{eq:p0x}
  p_0(x) = \Pi(x) \cos^2{(\delta
    +\alpha)} + \left(1-\Pi(x)\right) \cos^2\alpha, \\
  \label{eq:p1x}
  p_1(x) = \Pi(x)\sin^2{(\delta+ \alpha )} + \left(1-\Pi(x)\right)
  \sin^2\alpha,  
\end{gather}
where $\Pi(x)$ is defined by \refeq{eq:pi}.
The step sizes $\epsilon_i$, $i=0,1$, do not depend on $x$ and are given
by (see details in \ref{app:1}): 
\begin{gather}
  \label{eq:epsx00}
%%% see aux-14.05.13.nb
  \epsilon_0 = \operatorname{atanh}{\left(\frac{2\cos^2{(\alpha+\delta)}}{\cos^2{(\delta
    +\alpha)} + \cos^2\alpha}-1\right)} , \\
  \label{eq:epsx10}
  \epsilon_1 =
  \operatorname{atanh}{\left(\frac{2\sin^2{(\alpha+\delta)}}{\sin^2{(\delta
          +\alpha)} +\sin^2\alpha}-1\right)}.
\end{gather}

\refeqs{eq:p0x}{eq:epsx10} define obviously a Markovian random
walk. 

\subsection{Conditionally varied measurement parameters}

Suppose, we know exactly the initial state of the system $x_0$ and are
able to make all the rotations also exactly. In this case, the
subsequent positions $x_n$ of the qubit on $x$-line can be also
calculated exactly since we know the measurement outcomes $M_n=0,1$
and thus the step sizes $\epsilon_i$ at every $n$. We may now
introduce the state-dependent dynamics by allowing the parameters
$\delta$, $\alpha$ to be dependent on the step number $n$ and the
state of the system at the last step $x_n$. That is we may take some
(pre-defined) functions of two arguments $\alpha(n,x)$, $\delta(n,x)$
and at every step select the parameters for the next step
$\alpha_{n+1}$, $\delta_{n+1}$ as: $\alpha_{n+1} = \alpha(n,x_n)$,
$\delta_{n+1} = \delta(n,x_n)$.  In this way, the parameters $p_i$,
$\epsilon_i$ of our random walk are also some pre-defined functions of
$n$, $x_n$ defined by \refeqs{eq:p0x}{eq:epsx10}.

The functions $\alpha(m,x)$, $\delta(m,x)$ may be quite
arbitrary. They add new degrees of freedom to our system, leading, as
we will see, to rather interesting new dynamics.
We remark also that in quantum control schemes \cite{wiseman10:book}
the information about the current state of the system is often used by
feeding it back into the system via modification of the system's
Hamiltonian. In contrast, in our case, only the parameters of the
measurement itself, but not the parameters of system, are changed.

\subsection{Master equation}

Using \refeqs{eq:0}{eq:4} or \refeqs{eq:6}{eq:6a} it is easy to obtain
an equation governing the evolution of the probability density
function (pdf) $P(n,x)$, describing the probability $P$ of $\ket{q}$
to appear in the vicinity of $x$ at the step $n$. Since our qubit
$\ket{q}$ always remains in a pure state which is fully described by
its coordinate $x$ (or, equivalently, by $\theta$), such master
equation is just an another way to express the dynamics of
$\ket{q}$. It provides essentially the same information as
\refeqs{eq:0}{eq:4} or \refeqs{eq:6}{eq:6a}.  This reformulation will
be however useful in the next sections when we consider stochastic
ratchet behavior.

We start from the general case with no assumption about the particular
coordinate system.  We use the variable $y$ by which we may understand
any of the coordinates $x$, $\theta$ or $\Pi$ mentioned before.  We
introduce furthermore the measure $d\mathcal{P}(n,y) = P(n,y)dy$ which
expresses simply the total probability to find $\ket{q}$ in the
interval $[y,y+dy]$.  Then, by definition of our process, using the
Markov property and the formula for total probability
\cite{gardiner09:book} we obtain the following relation:
\begin{align}
\nonumber   
  d\mathcal{P}(n+1,y) = p_0(y_0(y))
  d\mathcal{P}(n,y_0(y)) + \\
  p_1(y_1(y)) d\mathcal{P}(n,y_1(y)),
  \label{eq:9}
\end{align}
where $y_i(y)$, $i=1,2$ are defined in an implicit way as
$y = y_i+\epsilon_i(y_i)$. This expression is valid for an arbitrary
(also varying) step size, that is, also for the state-conditioned
trajectories as they were defined above in the previous section.  In
the case of $x$-coordinates ($y\equiv x$) we obtain straightforwardly
the following expression for $P(n,x)$:
\begin{align}
\nonumber
    P(n+1,y) = p_0(x_0(x))
    x'_0(x)P(n,x_0(x)) + \\
  p_1(x_1(x)) x'_1(x) P(n,x_1(x)),
  \label{eq:10}
\end{align}
where $x = x_i(x)+\epsilon_i(x_i(x))$, $x'_i(x) = dx_i(x)/dx$.
In particular, for the constant measurement strength we have,
$\epsilon_i=\mathrm{const}$, $x'_i(x)=0$, and therefore we have:
\begin{align}
\nonumber
    P(n+1,x) = p_0(x-\epsilon_0)P(n,x-\epsilon_0) + \\
  p_1(x-\epsilon_1) P(n,x-\epsilon_1).
  \label{eq:me}
\end{align}

Very important are conserved quantities of \refeq{eq:9} or
\refeq{eq:10}.  The most obvious one is the average value of $\Pi$ on
$n$th step,
$\langle \Pi\rangle_n \equiv \int_{-\infty}^{+\infty} \Pi(x)
P(n,x)dx$, which represents the a priori probability to find $\ket{q}$
in the state $\ket{1}$ if we perform a projective measurement of
$\ket{q}$ after the $n$-th step of our process. One can show that from
\refeq{eq:me} it follows that:
\begin{equation}
%% see aux-03.04.13.nb for analytical proov.
  \label{eq:uppi_av}
  \langle \Pi\rangle_{n+1} = \langle \Pi\rangle_n,
\end{equation}
and thus for any $n$, $\langle \Pi\rangle_n = \langle
\Pi\rangle_0$. 
\refeq{eq:uppi_av} can be obtained by substituting \refeq{eq:10} into
definition of $\langle \Pi\rangle_{n+1}$, giving thus 
\begin{align}
\nonumber
& \langle \Pi\rangle_{n+1} =  \int_{-\infty}^{+\infty} \Pi(x) P(n+1,x)
dx = \\   \label{eq:av1a}
& \int_{-\infty}^{+\infty} P(n,x)\left\{p_0(x) \Pi(x) +
p_1(x) \Pi(x) \right\} dx,
\end{align}
where we made a replacement $x'_i(x)dx\to dx_i$ and the variable
change $x_i(x)\to x$ in both parts of the integral. Since
$p_0(x)+p_1(x)=1$, this gives \refeq{eq:uppi_av}.  We remark that
\refeq{eq:uppi_av} is universal, that is valid for any choice of the
measurement parameters, also if they vary in dependence on the step
$n$ or current position $x_n$.

\section{Continuous diffusive limit}
\label{sec:contin}

\subsection{general equation}

The continuous limit arises if we tend the measurement strength to
zero. In this case, instead of the discrete equation \refeq{eq:10}, a
continuous equation arises, with the step numbers $n$ being mapped to
a continuous ``time'' $t$.  If the measurement strength is constant
(independent on $n$) and if this constant strength tends to zero, the
corresponding limit is universal, that is, does not depend on the
measurement strength and on the particular measurement procedure.  The
dynamics in such ``unconditional'' continuous limit is often described
by the stochastic Schr\"odinger equation or by the master equation for
the density matrix
\cite{belavkin89,gisin84,wiseman96,brun02a,oreshkov05,varbanov07,wiseman10:book}.
Nevertheless, to our knowledge, a consideration general enough to
include time- and conditionally-varied measurements were presented
only very recently in \cite{bauer13}. Earlier works dealt only with
the case of measurements of equal strength or at least the strength
which is not explicitly time dependent (but might depend on time
indirectly via the outcome of the previous measurement)
\cite{oreshkov05,varbanov07}. Instead of directly writing the
resulting equation according \cite{bauer13} we will proceed, for the
sake of closeness of presentation, from the master equation for
$P(n,x)$, derived in the previous section, to the corresponding
continuous limit described by the Fokker-Planck (FP) equation.  The FP
approach used here is also different from \cite{bauer13} where Ito
calculus is used, but Ito and FP approaches are, of course, equivalent
\cite{gardiner09:book}. We use the later because of the
straightforward connection to the methods used in the theory of
stochastic ratchets \cite{reimann02}.

That is, our goal here is to derive the FP equation in the case which
includes the walk with conditionally varying measurement parameters %,
$\delta$, $\alpha$ which depend on the outcome of the all previous
measurements and also on $n$.
The transition to the continuous time can be done as follows: We
introduce ``time'' $t$ such that each step of our process corresponds
to a small interval
$\tau_n=\tau(\delta_n(n,x_{n-1}),\alpha_n(n,x_{n-1}))$, that is, we
replace $n=\sum^n_{i=1} 1$ by
\begin{equation}
t\equiv \sum_{i=1}^n\tau_i\label{eq:5}
\end{equation}
and allow $\tau_i(x_i)$ to tend to zero for every $i$, $x_i$. We do
not assume that all $\tau_i$ are equal.  In our case, as
$\tau_i(x_i)\to 0$, we can expect that
$P(t,x)\equiv \left.P(n,x)\right|_{n\to t}$ changes at every step only
slightly and we can then decompose $P(t,x)$ into series as:
\begin{equation}
P(t+\tau_n,x)\approxeq P(t,x) +
\tau_n \partial_tP(t,x).\label{eq:3}
\end{equation}

To be allowed to do this we must assume that, independently on $n$,
the step size
$\epsilon_{i,n}=\epsilon_i(\delta_n(n,x_{n-1}),\alpha_n(n,x_{n-1}))$
defined in \refeqs{eq:epsx00}{eq:epsx10} goes to zero as
$\tau_n\to 0$. In particular, this is the case if $\delta_n \to 0$,
$\alpha_n=\mathrm{const_n}>0$ for all $n$.  Thus, for small enough
$\delta_n$, we may assume:
\begin{gather}
\label{eq:alpha_n}
\alpha_n=\mathrm{const}(n,x),\\
\label{eq:delta}
  \delta_n=\delta g_\delta(x_{n-1},n),\\
  \tau_n=\delta^2 g_\tau(x_{n-1},n),
\label{eq:delta_tau}
\end{gather}
where we introduced the parameter $\delta\to0$ which describes how
fast $\delta_n$ and $\tau_n$ approach to zero; $g_{\tau}(x,n)>0$,
$g_{\delta}(x,n)$ are some functions which do not depend on $\delta$
and which we can chose at our will.

That is, we require that all $\tau_n$, $\delta_n$ tend to zero as
$O(\delta^2)$ and $O(\delta)$ respectively. This template is taken
from the consideration of the case with the constant step size as
shown in Appendix \ref{app:2}.  The functions $g_{\tau}(x,n)>0$ and
$g_{\delta}(x,n)$ provide ``form-factors'', which determine the
strength of measurement in dependence on the system position $x$ and
$n$.
Using Eqs.~(\ref{eq:5}),(\ref{eq:delta}),(\ref{eq:delta_tau}), we
define a 
function 
$g(x,t)$  as:
\begin{equation}
  \label{eq:25}
  g(x,t) =
  \left.\frac{g_\delta(x_{n-1},n)}{g_\tau(x_{n-1},n)}\right|_{n\to
    t;x_{n-1}\to x }.
\end{equation}
Using Eqs.~(\ref{eq:3}),(\ref{eq:25}) we derive, in a rather standard
way, the FP equation (see Appendix \ref{app:3} for details and a
description of the general procedure in \cite{gardiner09:book}):
\begin{gather}
  \label{eq:14}
  \partial_t P(t,x) = -\partial_x J(t,x),\\
J(x,y) = \mu(t,x) P(t,x)
  - \partial_{x}\left(D(t,x) P(t,x) \right).
  \label{eq:14a}
\end{gather}
%where 
Here
\begin{equation}
  \label{eq:mu:lim}
  \mu(x,t) = g(x,t)^2\tanh{(x)}, \, D(x,t)= g(x,t)^2/2,
\end{equation}
have now the meaning of the drift and diffusion coefficients,
respectively.  In different coordinates, like $\theta$ or $\Pi$ the FP
equation conserves its form, only the drift and diffusion coefficient
modifies (see Sec.~\ref{app:4}).

\begin{figure}[t!]%[htbp!]
  \begin{center}
    \includegraphics[width=0.5\textwidth]{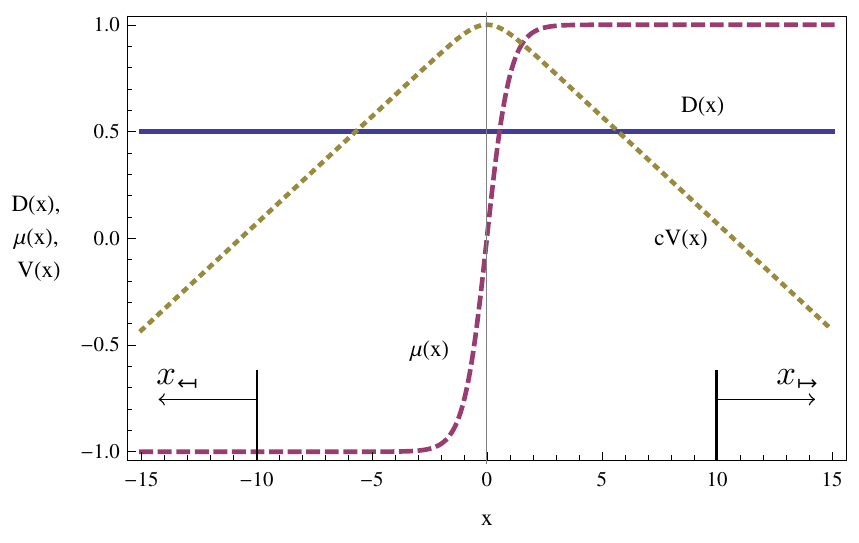}
    \caption{ \label{fig:diff}%%
      Diffusion $D(x)$ (solid blue line), drift coefficient $\mu(x)$
      (dashed red line) and the effective potential $V(x)$ (dotted
      yellow line, normalized to a constant $c=0.1$ for better
      visibility) in dependence on $x$ according to \refeq{eq:15}. The
      asymptotic coordinates $x_\lar$, $x_\rar$ defined in
      \refeqs{eq:xleft}{eq:xright} are shown, with $X=-10$ in this
      case. Asymptotic coordinates are useful when $|x|$ is large,
      that is, as the step $n\to\infty$: $D(x)$, $\mu(x)$ and $V(x)$
      are significantly simplified for large $|x|$.
 %%%
}
\end{center}
\end{figure}

This FP equation, as said, describes the dynamics of the pure state
$\ket{q}$ which position on the line between $\ket{0}$ and $\ket{1}$
is described by the coordinate $x$. Stochastic distribution of the
position $x$ is due to unpredictable character of the weak measurement
sequence.  The same FP equation describes also a Brownian heavily
damped particle moving in the potential
\begin{equation}
V(x,t) = -\int_0^x \mu(x',t)dx',\label{eq:30}
\end{equation}
 \cite{reimann02,doering98}. 
 The average drift velocity
 $\langle \dot{x}(t)\rangle \equiv \int_{-\infty}^{+\infty}
 \dfrac{dx}{dt} P(x,t) dx$ can be obtained also as an average of
 $J(x,t)$:
\begin{equation}
\langle \dot{x}(t)\rangle = \int_{-\infty}^{+\infty} J(x,t) dx.
\label{eq:31}
\end{equation}
Note that here the ensemble average is assumed, and
$\langle \dot{x}(t)\rangle$ depends on $t$.  For the case of $g(x)=1$,
that is, if the step size in our random walk is state-independent, we
have:
\begin{equation}
  \label{eq:15}
  \mu(x) = \tanh(x), \, V(x)=\ln{(\cos{(x)})},\, D=\frac12.
\end{equation}
The corresponding functions $\mu$, $D$, $V$ are shown in
\reffig{fig:diff}.
The solution of \refeqs{eq:14}{eq:14a},(\ref{eq:15}) wit the initial
condition $P(0,x) = \delta(x-X)$, where $\delta(x-X)$ is the Dirac
delta-function localized at the arbitrary point $X$ can be found
analytically \cite{gisin84}:
\begin{equation}
  \label{eq:21}
%% WBXIX:05.06.13, aux-05.06.13....nb
  P(t,x) = \frac{1}{\sqrt{2 \pi t}} \frac{\cosh x}{\cosh X}
  \exp{\left(-\frac{t^2+(x-X)^2}{2 t}\right)}. 
\end{equation}

\subsection{Asymptotic FP equation}

To make a semi-analytic approach described in \cite{reimann02} working
(as described in the next section), we have to find the conditions
where, assuming $g=\mathrm{const}$ in \refeq{eq:15}, we have also
$\mu=\mathrm{const}$.  In our equations this is generally not the case
because of $\tanh(x)$ factor. However, as one can see from
\reffig{fig:diff} and from \refeq{eq:15}, the deviation from this
condition decreases exponentially with $|x|$ because $|\tanh(x)|$
exponentially fast approaches 1.  Also, as one can see from
\refeq{eq:21}, if we take the initial starting point $X$ far away from
the origin $X=0$, $P(t,x)$ behaves very much like a normal Gaussian
distribution which shifts with time with the constant unit speed away
from $x=X$, and expands with the variance $\sigma^2=t$.

This allows us to consider the asymptotic behavior, as the initial
point $X$ and thus $x$ are far enough from the origin $x=0$.  We thus
introduce ``shifted'' coordinates $x_\lar$, $x_\rar$ as (see also
\reffig{fig:diff}):
\begin{gather}
\label{eq:xleft}
  x_\lar = x+X,\\
  x_\rar = x-X,
\label{eq:xright}
\end{gather}
where $X\gg0$ is a large arbitrary number. We will call them
``asymptotic coordinates''.  For such defined variables, neglecting
the terms which is exponentially small with $|X|$ we have from
\refeq{eq:mu:lim}:
\begin{gather}
  \label{eq:32}
  \mu(x_\lar,t) = -g(x_\lar,t)^2, \,
  \mu(x_\rar,t) = g(x_\rar,t)^2,\\
  \label{eq:32a}
  D(x_\lar,t) = g(x_\lar,t)^2/2, \,
  D(x_\rar,t) = g(x_\rar,t)^2/2,
\end{gather}
that is, the factor $\tanh{(x)}$ which were present  in the diffusion
coefficient in \refeq{eq:15}, disappears.
In the following, we will consider only the case when $x\to-\infty$,
and, correspondingly, we restrict ourselves to the variable $x_\lar$
(cf. \reffig{fig:diff}).  The dynamics for the case of $x\to+\infty$
is obviously analogous, only the overall drift direction will be the
opposite as \refeq{eq:32} indicates. The asymptotic FP equation for
this case coincides with the original one \refeqs{eq:14}{eq:14a}, only
written in asymptotic coordinates $x\to x_\lar$:
\begin{gather}
  \label{eq:14as}
  \partial_t P(t,x_\lar) = -\partial_x J(t,x_\lar),\\
J(x_\lar,y) = \mu(t,x_\lar) P(t,x_\lar)
  - \partial_{x_\lar}\left(D(t,x_\lar) P(t,x_\lar) \right).
  \label{eq:14asa}
\end{gather}
 
\subsection{FP equation for periodically varying potential}

In this section we focus on the case when $g(x_\lar,t)$ changes in
space and time periodically. We assume $g$ to have period $L$ in space
$x_\lar$.  In our new asymptotic coordinates, reformulation of the FP
equation \refeqs{eq:32}{eq:14asa} allowing to take advantage of such
periodicity is possible \cite{reimann02}. Namely, we define the
reduced quantities:
\begin{gather}
  \label{eq:ptilde}
\tilde{P}(x_\lar,t) = \sum_{j=-\infty}^{+\infty}
P(x_\lar+j L,t),\\
  \label{eq:jtilde}
\tilde{J}(x_\lar,t) = \sum_{j=-\infty}^{+\infty}
J(x_\lar+j L,t).
\end{gather} 
Obviously, $\tilde{P}(x_\lar,t)$ and $\tilde{J}(x_\lar,t)$ are finite
and defined in the range $x_\lar\in[-L/2,L/2]$.  Moreover, from
\refeqs{eq:ptilde}{eq:jtilde} one can see that $\tilde{P}$,
$\tilde{J}$ are periodic in $x_\lar$:
\begin{equation}
  \label{eq:34}
  \tilde{P}(x_\lar,t) = \tilde{P}(x_\lar+L,t),\,   \tilde{J}(x_\lar,t) = \tilde{J}(x_\lar+L,t).
\end{equation}

 \begin{figure}[tbph!]
  \begin{center}
    \includegraphics[width=\columnwidth]{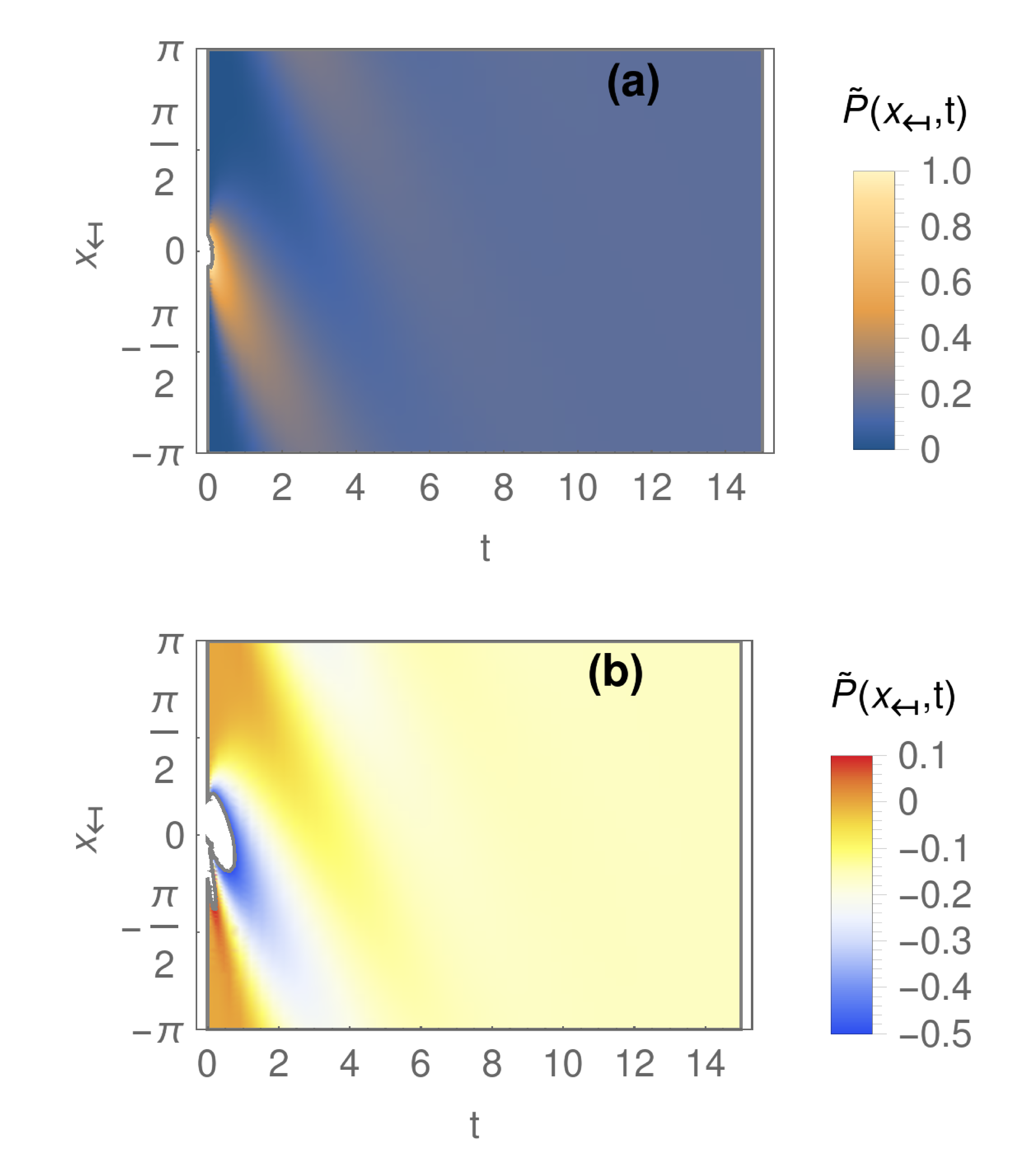}%%
    \caption{ \label{fig:prob-const-dist}%%
      The dynamics of the reduced asymptotic probability density
      $\tilde{P}(t,x_\lar)$ (a) and the current density
      $\tilde{J}(t,x_\lar)$ (b) for $g(t,x_\lar)=1$ obtained by direct
      simulations of \refeqs{eq:14b}{eq:14c}, assuming periodic
      boundary conditions and initial conditions described in text. In
      contrast to initial variables $P$, $J$, the reduced variable
      $\tilde{P}$, $\tilde{J}$ do have a steady-state, which is in
      this case a homogeneous distribution.}
  \end{center}
\end{figure}

In the asymptotic variables $\{x_\lar,t\}$, as it follows from
\refeqs{eq:32}{eq:32a}, $\mu(x_\lar,t)$ and $D(x_\lar,t)$ are periodic
in space with the same period $L$ (which is by the way not true for
$\mu$ and $D$ written using the original variable $x$).  Under these
circumstances the FP equation written for $\tilde{P}$, $\tilde{J}$
remains the same as for $P$, $J$.  That is, we have:
\begin{gather}
  \label{eq:14b}
  \partial_t \tilde{P}(t,x_\lar) = -\partial_{x_\lar} \tilde{J}(t,x_\lar),\\
\tilde{J}(x_\lar,t) = \mu(t,x_\lar) \tilde{P}(t,x_\lar)
  - \partial_{x_\lar}\left(D(t,x_\lar) \tilde{P}(t,x_\lar) \right) ,
  \label{eq:14c}
\end{gather}
where the coefficients remain the same as before, that is, are given by \refeqs{eq:32}{eq:32a}. 
The advantage of such reformulation is that now we can consider only
the finite interval in $x_\lar$ from, say, $-L/2$ to $L/2$. Besides,
the equation for the average drift velocity \refeq{eq:31} also retains
its form:
\begin{equation}
  \label{eq:33}
  \langle \dot{x}_\lar(t)\rangle = \int_{-L/2}^{L/2} \tilde{J}(x_\lar,t) dx.
\end{equation}
Remarkably, the direct definition of $\langle \dot{x}_\lar(t)\rangle$ as
the average of $\dot{x}_\lar$ with the probability distribution
$\tilde P(x_\lar,t)$ is not valid anymore.

As an illustration of the dynamics appearing in the reduced equations,
we show in \reffig{fig:prob-const-dist} the dynamics of
$\tilde{P}(x_\lar,t)$, $\tilde{J}(x_\lar,t)$ for the case of
$g(x_\lar,t) = \mathrm{const} = 1$ obtained using direct numerical
simulations of \refeqs{eq:14b}{eq:14c} with the initial condition
$P(x_\lar,t)\propto \exp{(-x_\lar^2/0.1)}$ and periodic boundary
conditions.  The figure shows rather rapid homogenization of
$\tilde{P}(x_\lar,t)$, $\tilde{J}(x_\lar,t)$ in space because of the
action of diffusion.
This homogenization illustrates an important peculiarity of the
reduced quantities: although the initial variables $J$, $P$ have no
steady-state in their dynamics, the reduced quantities $\tilde J$,
$\tilde P$ do have a steady-state. In the case of
\reffig{fig:prob-const-dist} this steady state is simply a constant
which does not depend neither on $t$ nor on $x_\lar$.
 
 \begin{figure}[tbph]
  \begin{center}
    \includegraphics[width=\columnwidth]{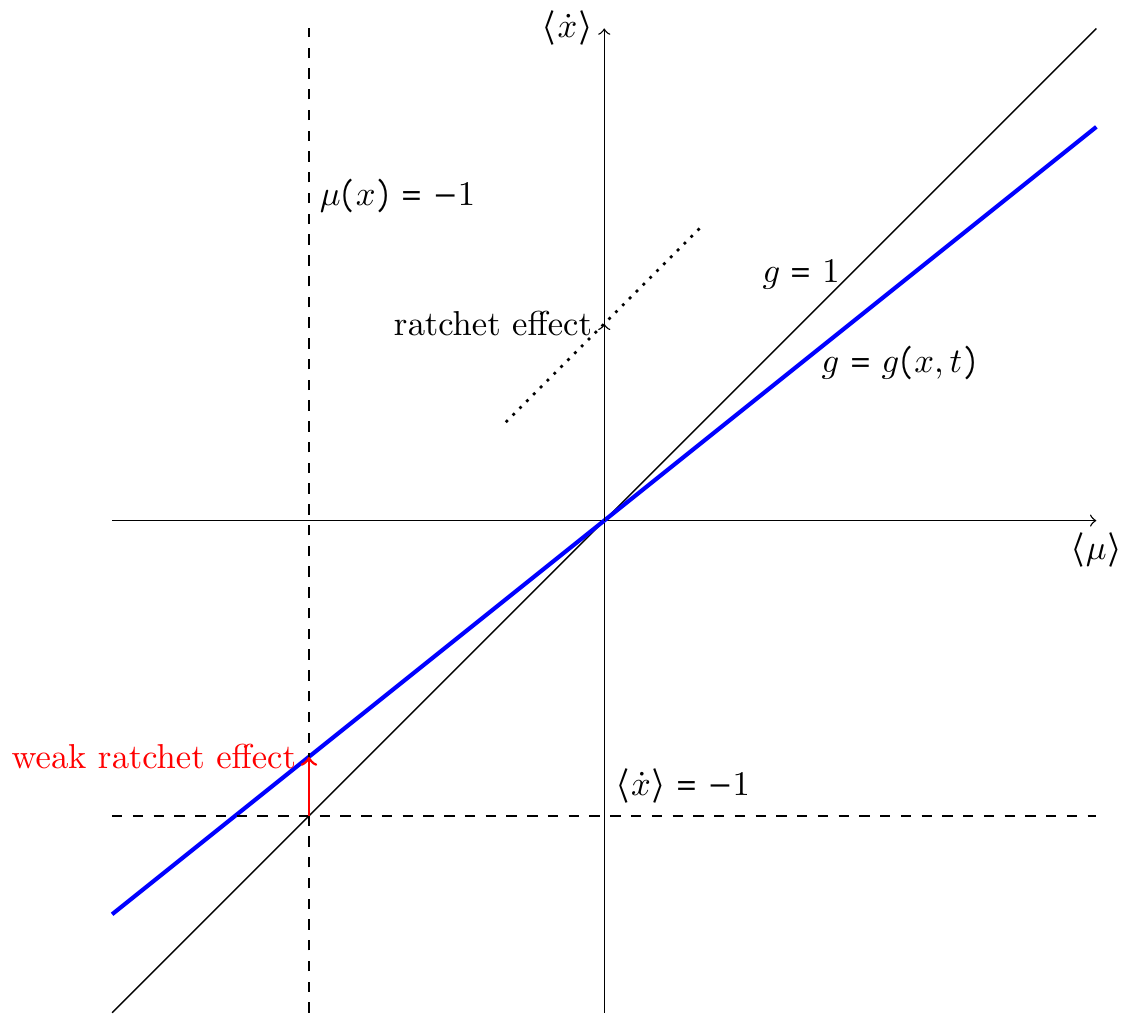}%%
    \caption{ \label{fig:ratchet}%%
      Schematic representation of the Brownian ratchet effect. Without
      a ratchet effect ($g=1$), a constant drift force $\mu$ leads to
      a current $\langle \dot x \rangle=\mu$ (black thin line). In
      contrast, when $g$ changes in space and/or time (but still
      $\langle g\rangle=1$), the average current
      $\langle \dot x \rangle$ can be modified even through the
      average force $\langle \mu \rangle$ remains the same (blue solid
      and black dotted lines). Although in many other systems the
      Brownian ratchet effect can change the average direction of
      motion (see black dotted line), it is not possible in the
      present case -- the type of ratchet which we call ``weak
      ratchet''. In this later case, $|\langle \dot x \rangle|$ can be
      modified but its sign is not reversed (see blue line). The
      asymptotic values of $\mu$ and $\langle \dot x \rangle$ for
      $x\to-\infty$ (that is, assuming asymptotic coordinates
      $x_\lar$) are marked by dashed lines. Weak ratchet effect in the
      asymptotic case is marked by a red arrow.}
  \end{center} 
\end{figure}

\section{Brownian ratchets}
\label{sec:ratchets}

 \begin{figure*}[tbph]
  \begin{center}
    \includegraphics[width=0.75\textwidth]{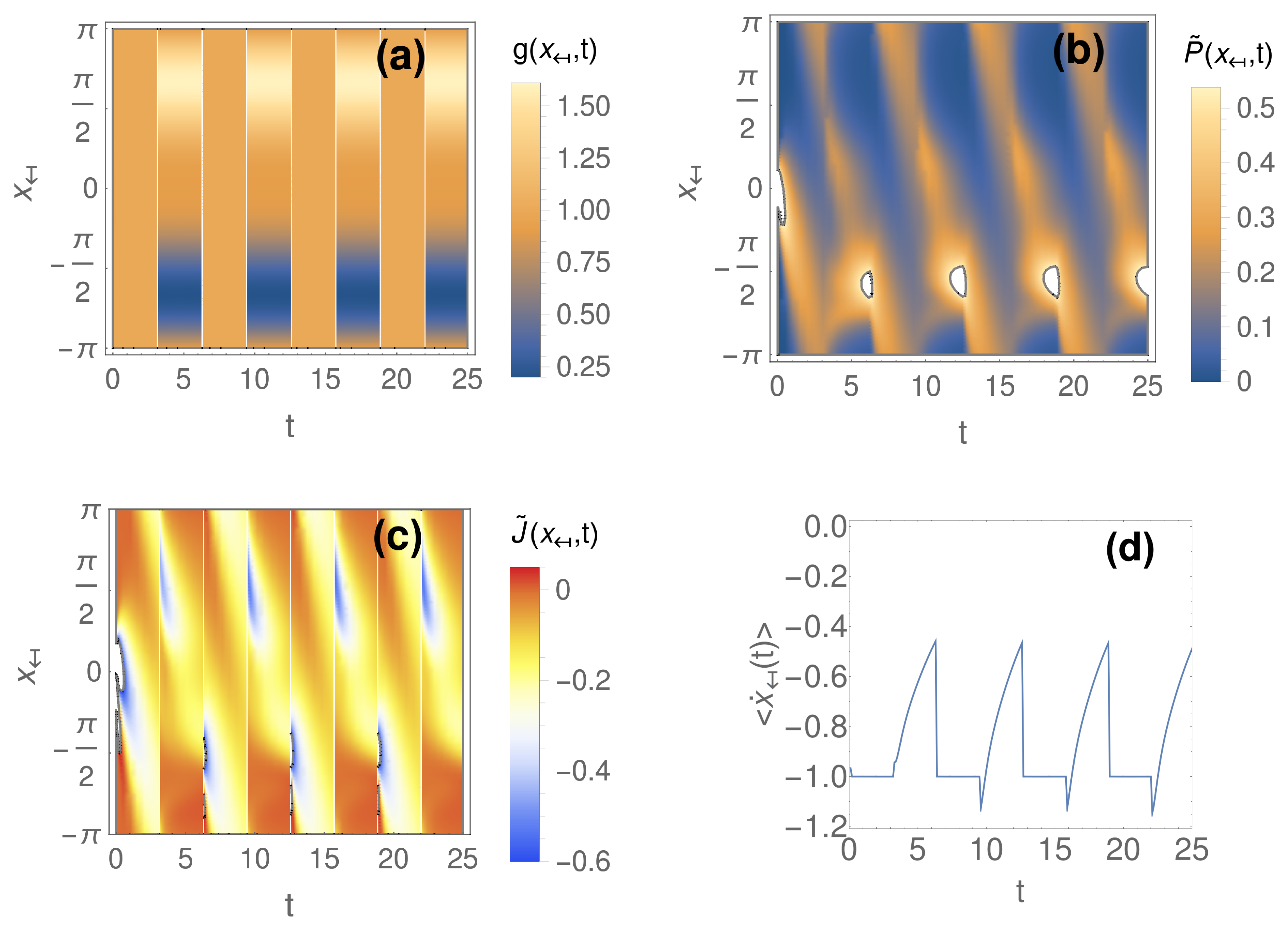}%%
    \caption{ \label{fig:spt-ratchet}%%
      Brownian ratchet effect for $g(x_\lar,t)$ varying in space and
      time. (a) $g(x_\lar,t)$, as given by
      \refeqs{eq:37}{eq:22b}. (b), (c) the reduced asymptotic
      probability density $\tilde{P}(x_\lar)$ and the current density
      $\tilde{J}(x_\lar)$ obtained by direct simulations of
      \refeqs{eq:14b}{eq:14c} with periodic boundary conditions and
      initial conditions described in text. (d) the averaged current
      $\langle \dot x_\lar(t)\rangle$ in dependence on time.}
  \end{center}
\end{figure*}

One of the most interesting phenomena in Brownian flows is a
possibility of so called stochastic ratchets
\cite{haenggi09,reimann02}. Namely, by manipulating dynamically the
potential $V(x,t)$ in a Brownian flow, one can have nonzero average
motion $\langle \dot{x}\rangle\ne0$ even in the case when the average
force, $\langle\mu\rangle\equiv\int \mu(x,t)dx$ is exactly zero (for
every $t$). Here, to simplify notations, we used the denotation
$\langle \dot{x}\rangle$ for the time- and space average defined as:
\begin{equation}
\label{eq:avxt}
\langle \dot{x}\rangle \equiv \langle \bar{\dot{x}}(\infty) \rangle,  
\end{equation}
where:
\begin{equation}
  \label{eq:23}
  \langle \bar{\dot{x}}(t)\rangle \equiv 
  %\lim_{t\to\infty}
\frac1t\int_0^t\langle \dot{x}(\tau)\rangle\,d\tau
\end{equation}
is the ''moving average'' in time of the space average.
Alternatively, one can speak about a ratchet effect if a nonzero
initial force $\mu\ne0$ can be canceled or even reversed by
introducing some periodic modulations of the potential.  Both of these
definitions are visualized in \reffig{fig:ratchet}.

In our case it is quite clear that the average flow defined by
$\mu=-1$ (in the asymptotic case $x\to x_\lar$) can not be
reversed. Otherwise, one would have a possibility to violate the
conservation law of $\langle\Pi\rangle$ given by \refeq{eq:uppi_av} by
tuning, for every particular trajectory, the potential in such a way
that the current system state is forced to move in the direction
opposite to $\mu$ and thus bring our system to any of the states
$\ket{0}$, $\ket{1}$ at our wish which would violate
\refeq{eq:uppi_av}.

Nevertheless, one can try to find a Brownian ratchet effect in a weak
sense, that is, to find such a function $g(x_\lar,t)$ that the the
asymptotic value of $\langle\dot x_\lar\rangle>-1$, despite of
$\langle\mu\rangle=-1$.  The notion of a weak ratchet effect, in
comparison to a ``normal'' stochastic ratchet, is visualized in
\reffig{fig:ratchet} (blue line).  Weak ratchets are in close
correspondence to the weak Parrondo games, where the combination of
lossy games lead to less lossy one, but still not to a winning one
\cite{wu14}.
Of course, one can always obtain $\mu(x_\lar,t)=0$ by simply putting
$g=0$, that is, by reducing step size of the random walk to zero.
Here we want however to investigate the effects which is independent
on such raw step size reduction.  To ensure this we will always take
such $g$ that
\begin{equation}
  \label{eq:35}
  \langle g(t)^2\rangle  = 1,
\end{equation}
where we define $\langle g(x_\lar,t)^2\rangle$ as:
\begin{equation}
  \label{eq:36}
  \langle g(t)^2\rangle \equiv
\frac{1}{L}\int_{-L/2}^{L/2} g(x_\lar,t)^2
  dx_\lar.
\end{equation}
This condition excludes the possibility to reduce
$\langle\dot x_\lar\rangle$ by reducing the measurement strength
globally. That is, if one reduces the measurement strength near some
point, one has to increase it in the vicinity of some another one.

We will now try to construct a periodic in time and space function $g$
which allows to reduce $|\langle \dot x_\lar \rangle|$, making it as
small as possible.

We remark that several various types of stochastic ratchets has been
considered in the literature (see \cite{reimann02,braun:book04} and
references therein), the classification is based on the functional
form of $D$ and $\mu$. In many commonly studied hydrodynamic Brownian
flows $\mu$ and $D$ can be varied quite independently -- in contrast
to our situation where independent variation of $D$ and $\mu$ is
impossible because of the common factor $g$.  Our situation closely
resembles hydrodynamic Brownian ratchets with varying friction
\cite{reimann02,luchsinger00,lancon01,krishnan92}.  The most studied
class of ratchets is so called pulsating ones, where $\mu$ may vary in
space and time, whereas $D$ is a constant. On the other hand, the
situations when both $\mu$ and $D$ vary in space or, alternatively, in
time, were also considered under the names Seebeck or temperature
ratchets, correspondingly. They can be mapped, by suitable change of
variable, to the pulsating ratchets.

In our case, as one can see, the situation when $g$ is changing in
time but not in space provides no possibility for any ratchet
effect. Namely, in this case \refeq{eq:33} can be calculated directly
by integrating \refeq{eq:14c} with the boundary conditions
\refeq{eq:34}, giving $\langle \dot x_\lar(t)\rangle=-1$.  We have
then $\tilde P\to\mathrm{const}=1/L$, that is, full homogenization of
$\tilde P$ will take place, exactly as in the case of $g=1$.

We can therefore consider the cases when $\mu$ and $D$ change both in
time and space or only in space. For the presence of the ratchet
effect, the symmetry of the $\mu$ and $D$ are of the critical
importance. In general, ``almost all'' functions except the ones
processing certain particular symmetry properties allow the ratchet
effect \cite{reimann02,braun:book04}. Nevertheless, no analytical
symmetry relation is known, to our knowledge, for the case when both
$\mu$ and $D$ are arbitrary functions of space and time. For the case
when $\mu$ and $D$ are only space-dependent, the situation is simpler
and is discussed in the next section.

One of the most well-known types of ratchets is an on-off tilting
ratchet, were the diffusion $D$ is constant and the asymmetric
potential $V$ is switched on and off. At the on-stage, the particle
moves to the minimum of the potential and therefore becomes well
localized. When the potential is switched off, diffusion leads to a
broadening of the particle's wave packet. Switching the potential on
again makes the particle feeling the force, which pushes it to certain
direction. If the potential is asymmetric, the force is also
asymmetric, leading to an average current.

Having in mind said above, we first probe functions $g(x_\lar,t)$
which has the following form:
\begin{equation}
\label{eq:37}
g(x_\lar,t)=C(t)\left\{1+F(x_\lar)f(t)\right\},
\end{equation}
where the function $f(t)$ is periodic in time which models the
switching on and off behavior, and $F(x)$ is periodic in space, but
might be asymmetric. The normalizing constant $C(t)$ is obtained from
\refeq{eq:35}.  To start with, we will try the following functions:
  \begin{gather}
    \label{eq:22}
    f(t) = (\sign{(\sin{(t)})}-1)/2, \\
    F(x_\lar) = a\left[\sin{(x_\lar)}+b\sin{(2
      x_\lar)}\right], \label{eq:22a}\\
  a=-0.6,\: b=-0.5.\label{eq:22b}
\end{gather}
Here, $f(t)$ works as a switcher which is active only half of the
period, $a$ determines the ``amplitude'' of the periodic potential
whereas $b$ is selected in such a way that the shape of $g$ resembles
a ''saw-tooth'' one, in order to introduce some spatial asymmetry into
the profile $g$. Indeed, this shape of $F$ is simply the decomposition
of the ideal saw-tooth shape
$F_s(x)=\sum_{n=1}^{\infty}(-1)^n\sin{n x}/n$ which is cut off on the
second term.

\begin{figure}[tbph!]
    \includegraphics[width=\columnwidth]{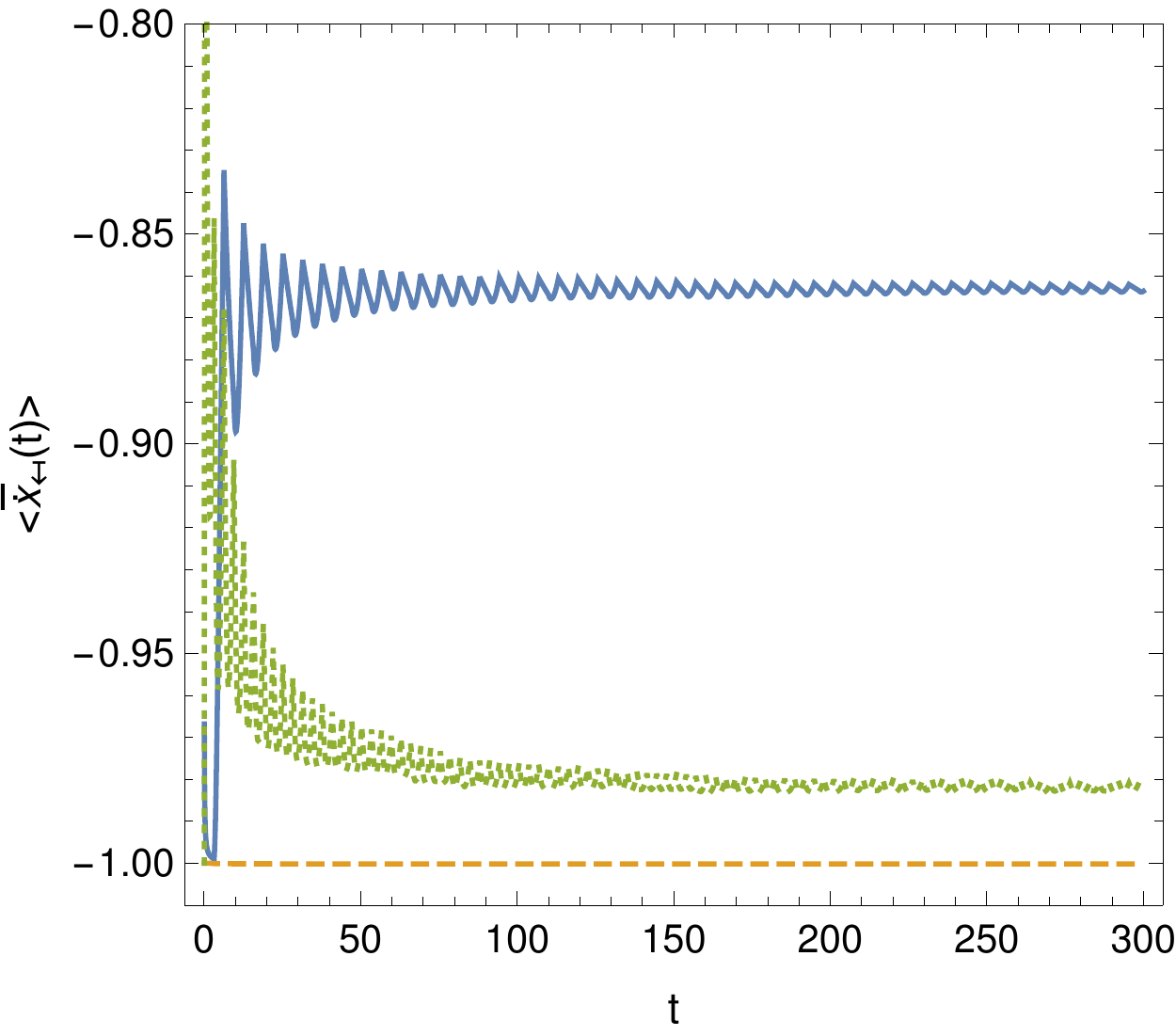}%%
    \caption{ \label{fig:ratchet-effect}%%
      Dependence of the temporal average
      $\langle \bar{ \dot{ x}}_\lar\rangle$ given by \refeq{eq:23} on
      the averaging time interval $t$. In the case of constant $g=1$
      (orange dashed line) this quantity quickly approaches $-1$ (no
      ratchet effect), whereas in the case of varying measurement
      strength with the parameters \refeqs{eq:37}{eq:22b} (blue solid
      line) it approaches $\approx -0.86$, demonstrating a weak
      Brownian ratchet. The effect strength depends on the function
      shape. For instance, green dotted line shows the case
      \refeq{eq:37} with the spatial dependence $F(x)$ given by
      \refeq{eq:22a} with $a=-0.8$, $b=0$, and $f(t)=\sign{\sin{t}}$.
    }
\end{figure}

The resulting dynamics is plotted in \reffig{fig:spt-ratchet}. Namely,
the shape of $g$ is presented in \reffig{fig:spt-ratchet}(a) whereas
\reffig{fig:spt-ratchet}(b) and \reffig{fig:spt-ratchet}(c) show the
temporally- and spatially resolved probability and current. To
calculate \reffig{fig:spt-ratchet}, the boundary and initial
conditions were taken as in \reffig{fig:prob-const-dist}.
One can see from \reffig{fig:spt-ratchet}(b) that when the
space-varying potential is on, the probability density $\tilde P$
concentrates in the regions where $g$ (and thus $D$, $\mu$) is
minimal. When it is off, the wave packet starts to diverge (and at the
same time is moving to the negative direction of $x_\lar$).
This behavior is thus different from typical pulsating ratchets in the
sense that when the potential is switched on, the particle is
localized in the minima of $g$ (and thus of $D$ and $\mu$ and not in
the minima of the potential.  In \reffig{fig:spt-ratchet}(d) one can
see that the current $\langle \dot x_\lar(t)\rangle$ approaches, after
a short transition process, a stationary regime of oscillations in
time with a period $2\pi$.  The long time behavior of the average of
$\langle \bar{ \dot x}_\lar(t)\rangle$ given by \refeq{eq:23} is shown
in \reffig{fig:ratchet-effect}, where it is seen that this average
approaches $\approx -0.86$ instead of $-1$ as in the case of constant
$g=1$, $\mu=-1$, thus clearly showing the ratchet effect in this
process.

To check the stability of the effect, simulations were made for
different functions $f(t)$, $F(x)$. For instance, in
\reffig{fig:ratchet-effect} the case with $b=0$ and
$f(t)=\sign{(\sin{(t)})}$ is also plotted. In this case, the ratchet
effect is definitely smaller but still persists.  As said, the
condition \refeq{eq:35} excludes the effect of bare step size
reduction in this random walk, demonstrating that the ratchet effect
is a dynamical phenomenon independent from the step size.

\section{Seebeck ratchets and dynamical localization}
\label{sec:localiz}

 \begin{figure*}[tbph!]
  \begin{center} 
    \includegraphics[width=0.75\textwidth]{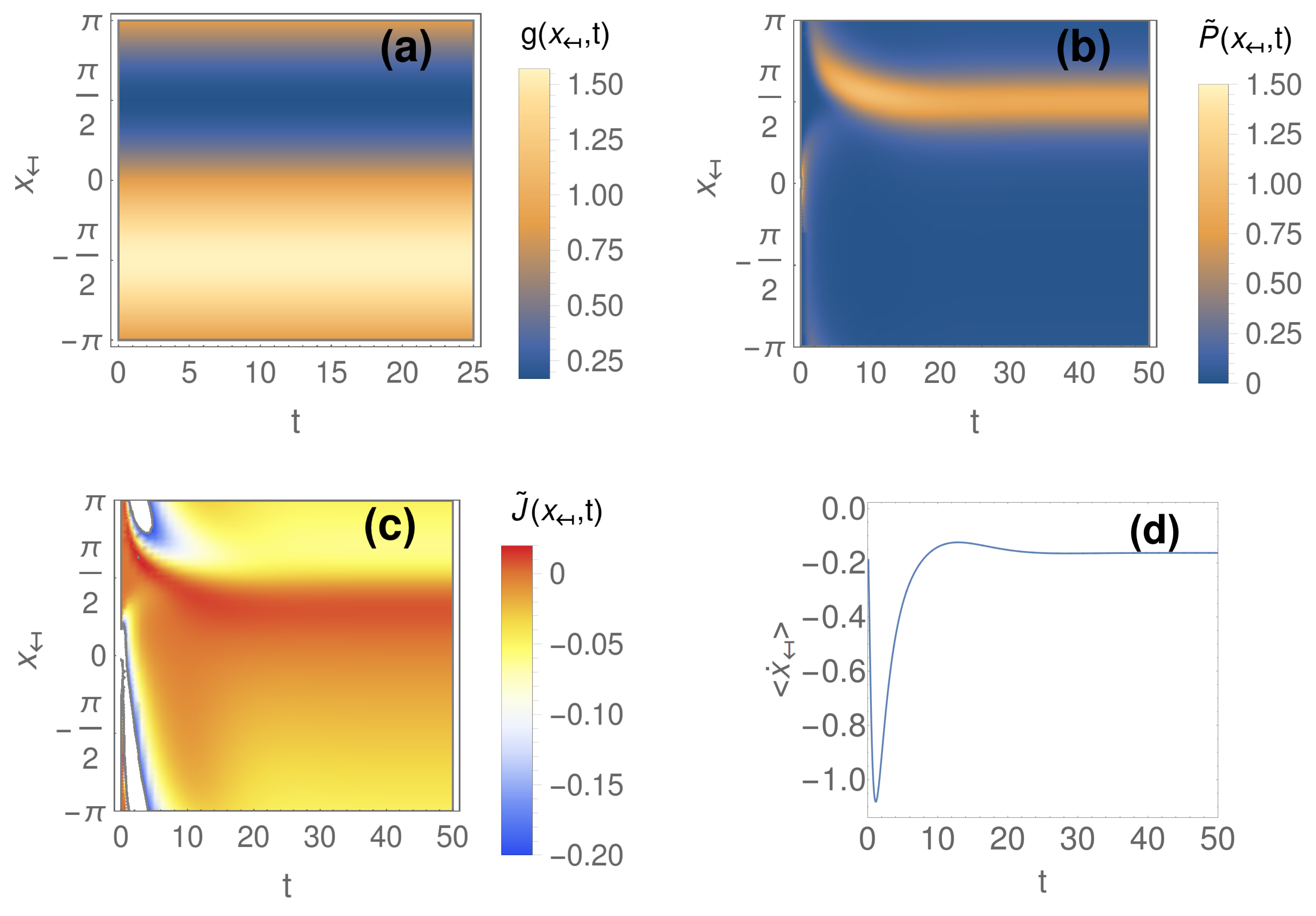}%%
    \caption{ \label{fig:sp-ratchet}%%
      Seebeck ratchet and dynamical localization effect for
      $g(x_\lar)$ dependent only on spatial coordinate. (a)
      $g(x_\lar,t)$ given by \refeq{eq:37} with $C(t)=\mathrm{const}$,
      $f(t)=1$, and other parameters defined by \refeq{eq:22a},
      \refeq{eq:22bb}.  (b), (c) Reduced asymptotic probability
      density $\tilde{P}(x_\lar)$ and the current density
      $\tilde{J}(x_\lar)$ obtained by direct simulations of
      \refeqs{eq:14b}{eq:14c} with periodic boundary conditions and
      initial conditions described in text. (d) Spatially averaged
      current $\langle \dot x_\lar\rangle$ in dependence on time.  }
  \end{center}
\end{figure*}

\begin{figure}[t!]%[htbp!]
  \begin{center}
    \includegraphics[width=\columnwidth]{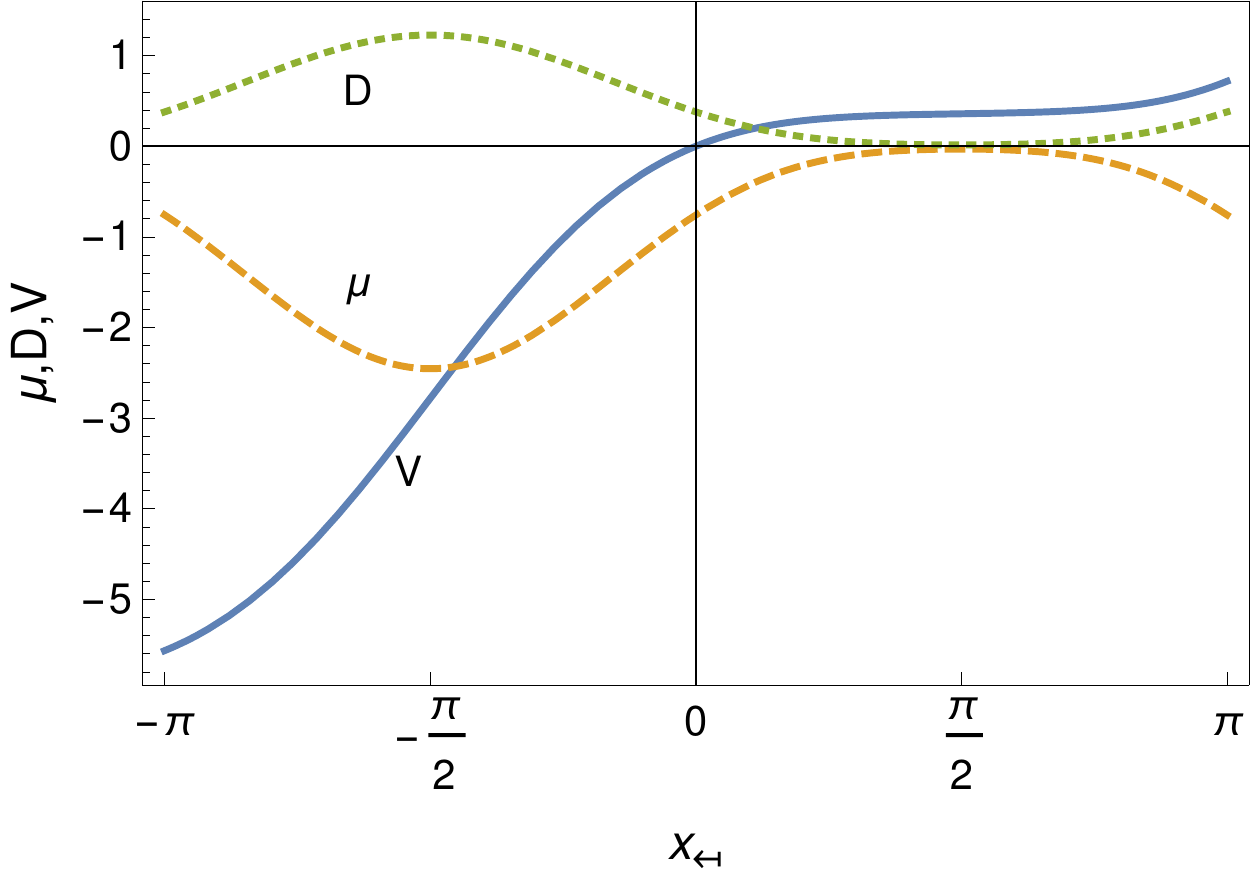}%%
    \caption{ \label{fig:V-sp}%%
      Diffusion $D(x)$ (solid blue line), shift $\mu(x)$ (dashed red
      line) and the effective potential $V(x)$ (dotted yellow line,
      normalized to a constant $c=0.1$ for better visibility) in
      dependence on $x_\lar$ with $g(x_\lar)$ being time-independent, that is, given by \refeq{eq:37} with $C(t)=\mathrm{const}$, $f(t)=1$, and other parameters defined by \refeq{eq:22a}, \refeq{eq:22bb}. 
 %%%
    }
\end{center}
\end{figure}

In general, the ratchet effect can appear if $D$, $\mu$ change only in
space. In this case, one may have the diffusion $D$ and the potential
$V$ defined by \refeq{eq:30} being not in phase \cite{reimann02},
which is in our case is typically fulfilled, since $D\sim\mu$,
$\mu=-\partial_x V$ (that is, if $D\sim\sin x$, then $V\sim\cos x$;
see also \reffig{fig:V-sp}). Such ratchets are typically known as
Seebeck ones \cite{reimann02}.  For Seebeck ratchets, an analytical
condition exists which determines the absence of the ratchet
effect. In particular, if we consider the case with no average force
($\langle \mu \rangle=0$) and if $\int_{-L/2}^{L/2}\mu(x)/D(x)dx=0$,
no ratchet effect is present \cite{vankampen1988,landauer88}.  In our
case, $\langle \mu \rangle\ne0$ so that the condition above can not be
applied directly. Nevertheless, we can, by the replacement
$x_\lar\to x_\lar$, $t\to t-x_\lar$, reduce our equation to the case
with $\langle \mu \rangle=0$. In this case we have
$\langle \dot x_\lar\rangle\to \langle \dot x_\lar\rangle +1$.
Afterwards, we can apply the above criterion, which gives us the
criterion for the absence of the ratchet effect for our case in the
form:
\begin{equation}
\label{eq:seebeck_cr}
\frac{1}{L}\int_{-L/2}^{L/2}\frac{1}{g^2(x_\lar)}dx_\lar =1.
\end{equation}
That is, for a typical function $g$ (which satisfies \refeq{eq:35}) we
should expect the presence of a ratchet effect, unless
\refeq{eq:seebeck_cr} is also valid.  An exemplary profile of $g$
which we use to test the Seebeck ratchet numerically is given by
\refeq{eq:37} with $f(t)=1$ and $F(x_\lar)$ defined by \refeq{eq:22a}
with:
\begin{equation}
  a=-0.8,\: b=0,\label{eq:22bb}
\end{equation}
and is shown in \reffig{fig:sp-ratchet}(a).  For such a function $g$,
as one can see in \reffig{fig:sp-ratchet}(b), the average current
$\langle \dot x_\lar\rangle$ can be also larger than $-1$; in the case
of \reffig{fig:sp-ratchet} it approaches $\approx -0.2$ as one can see
in \reffig{fig:sp-ratchet}(d).  In this case, the initial distribution
is quickly rearranged to a stationary (but inhomogeneous) one.

Now, again, the system is located mostly near the minimum of $g$. This
allows interpretation of the Seebeck ratchet effect in the present
case in the terms of a ''dynamical localization''.  Namely, let us
observe the potential $V(x_\lar)$ as shown in \reffig{fig:V-sp} (solid
blue line). One can see that $V(x_\lar)$ approaches a flat region
(where it is almost constant) close to $x_\lar=\pi/2$. That is, there
is almost no effective force at that point. If our effective
``particle'' approaches this region, it nearly stops. Nevertheless,
the ``particle'' experiences some small drift to the negative
direction of $x_\lar$.

\begin{figure}[htbp!]
  \begin{center}
    \includegraphics[width=0.4\textwidth]{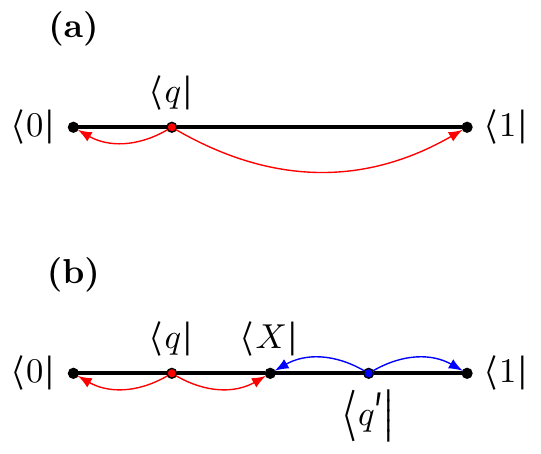}%%
    \caption{ \label{fig:pm}%%
      The limits $t\to\infty$ of the weak measurement sequence  in
      the case of $g=1$ (a) and in the case of $g(x,t)$
      such that $g(x)\to0$ as $x\to X$ (b); the coordinate $X$
      corresponds to the qubit state $\ket{X}$. In the former case,
       an arbitrary state
      $\ket{q}$ approaches either to $\ket{0}$ or $\ket{1}$, whereas in the case
      (b) the  state may also have
      $\ket{X}$ as a limiting point. Some states
      (such as $\ket{q}$) tend to either $\ket{0}$ or $\ket{X}$,
      the others (as $\ket{q'}$) approach $\ket{1}$ or $\ket{X}$.
 %%%
    }
\end{center}
\end{figure}

Going one step further, we consider now the case when $g=0$ at some
point. In this case we also expect localization shown in the previous
example. But more interesting dynamics will also appear as we will see
below. In general, to observe localization, it is not necessary to
take the periodic potential as it was in the previous example.  We now
consider the global dynamics related to localization, and therefore we
return back from ``asymptotic coordinate'' $x_\lar$ to the initial
coordinate $x$ and thus to the FP equation as written in
\refeqs{eq:14}{eq:14a}. We assume also for simplicity that $g(x)$
approaches zero only in one single point $X$, that is, $g\to 0$ as
$x\to X$.  Returning to our initial qubit, the state $\ket{X}$ is
given by
\begin{gather}
\label{eq:41}
\ket{X}=A\ket{0}+B\ket{1};\\
A=\sqrt{\Pi(X)},
\,B=\sqrt{1-\Pi(X)},\label{eq:41a}
\end{gather}
where $\Pi(X)$ is given by \refeq{eq:pi}.  As we will see later the
trajectory can not cross $\ket{X}$ in this case.  A state $\ket{x}$
located between of $\ket{0}$ and $\ket{X}$ will approach either
$\ket{0}$ or $\ket{X}$ as $t\to\infty$. Analogously, a state located
initially between $\ket{X}$ and $\ket{1}$ will approach either
$\ket{X}$ or $\ket{1}$ (see \reffig{fig:pm}(b)).  We note a similarity
to the initial system with the state-independent coupling strength
$g=1$ in this limiting dynamics (where the limiting states are
$\ket{0}$ and $\ket{1}$, see \reffig{fig:pm}a).  One can make this
analogy exact by considering the FP equation in coordinates $\Pi(x)$
defined in \refeq{eq:pi} which is given by (see also
Sec.~\ref{app:4}):
\begin{gather}
  \label{eq:16aa}
  \partial_t P(t,\Pi) =  \partial_{\Pi\Pi}\left(D(\Pi)P(t,\Pi)\right)\\
 D(\Pi) = 2(\Pi-1)^2\Pi^2g^2(\Pi), \label{eq:16aaa}
\end{gather}
%%%
so that the diffusion coefficient $\mu=0$ in these
  coordinates. We remark that for $\ket{0}$ and $\ket{1}$
($\Pi=0$ and $\Pi=1$ correspondingly) $D(\Pi)=0$.

Let us make the denotation $\Pi(X)\equiv \Pi_X$; we have thus
$g(\Pi_X)=0$. We also consider only one case when the initial state is
in between $\ket{0}$ and $\ket{X}$, that is, $x(t=0)<X$ and
$\Pi(t=0)<\Pi_X$ (see \reffig{fig:pm}(b), red lines). In this case we
can obviously define such function $\tilde g(\Pi)$ that:
%% WBA4:18.10.16, WBA5:26.01.17
\begin{equation}
  \label{eq:7}
  g(\Pi) = \frac{\Pi_X-\Pi}{1-\Pi}\tilde g(\Pi),
\end{equation}
which is possible without singularities since $1-\Pi>1-\Pi_X>0$. 
Here, $\tilde g\ge 0$ does not anymore necessarily approaches to zero as
$\Pi\to\Pi_X$.
Now, by making a variable change:
\begin{equation}
\tilde\Pi=\Pi/\Pi_X,\, \tilde t = \Pi^2_Xt,\label{eq:8}
\end{equation}
we arrive to a new FP equation:
\begin{gather}
  \label{eq:16bb}
  \partial_{\tilde t} P(\tilde t,\tilde \Pi)
  =  \partial_{\tilde\Pi\tilde\Pi}\left(\tilde D(\tilde\Pi)P(\tilde t,\tilde\Pi)\right),\\
 \tilde D(\tilde\Pi) = 2(\tilde\Pi-1)^2\tilde\Pi^2\tilde g^2(\tilde\Pi), \label{eq:16bbb}
\end{gather}
where $\tilde g(\tilde\Pi)=\tilde g(\Pi \Pi_X)$.  One can see that
\refeqs{eq:16aa}{eq:16aaa} and \refeqs{eq:16bb}{eq:16bbb} are
completely equivalent. That is, the dynamics of the random walk
between $\ket{0}$ and $\ket{X}$ and between $\ket{0}$ and $\ket{1}$
can be one-to-one-mapped to each other. In particular, the dynamics of
the random walk with $\tilde g=1$, that is, with
$g(\Pi) = (\Pi_X-\Pi)/(1-\Pi)$, is completely equivalent to the
dynamics of the simplest random walk with $g=1$.  The system with
$\tilde g=1$ behaves near $\ket{X}$ in the same way as the system with
$g=1$ near the state $\ket{1}$, for instance, the time of arrival to
the point $\ket{X}$ is infinite. This is obviously true for any other
bounded functions $\tilde g$ obeying \refeq{eq:35} and such that $g>0$.

This allows also to calculate straightforwardly the probability of the
outcomes $\ket{0}$ or $\ket{X}$ (resp. $\ket{X}$ or $\ket{1}$) as
$t\to\infty$. In our initial system with $g=1$ the a priori
probabilities to have $\ket{0}$ or $\ket{1}$ would be given by
$\Pi(x)$ and $1-\Pi(x)$, correspondingly, and, according to
\refeq{eq:uppi_av}, also do not depend on the measurement strength
$g(x,t)$ (unless $g$ approaches zero somewhere). By rescaling the
latter situation using \refeq{eq:8} we see, that, starting from the
state $\ket{x}$ we reach $\ket{0}$ or $\ket{X}$ (resp.  $\ket{X}$ and
$\ket{1}$) with the probabilities $\tilde \Pi(x)=\Pi(x)/\Pi_X$ and
$1-\tilde \Pi(x)=1-\Pi(x)/\Pi_X$. This probability also does not
depend on the measurement strength (unless it approaches zero
somewhere else at $x<X$). In the same way, if $\ket{x}$ is in between
$\ket{X}$ and $\ket{1}$, we obtain that the probabilities to reach
$\ket{X}$ or $\ket{1}$ are $(\Pi(x)-\Pi_X)/(1-\Pi_X)$ and
$(1-\Pi(x))/(1-\Pi_X)$ respectively.

\section{Conclusions}
\label{sec:concl}

In the present article we have considered quantum trajectories
resulting from a sequence of weak measurements, in the simplest
one-dimensional settings, but assuming the measurement strength
depending on the step number $n$ and on the current state of the
system described by the coordinate $x$ on the line.  Of course, the
current state can not be inferred from the measurement directly, in
contrast to classical systems. Nevertheless, if the initial state and
the parameters of the weak measurements are known, all the subsequent
positions of the system can be inferred from the sequence of the
measurement outcomes, and thus the conditional measurement strength
can be well defined.

Such measurement process, in the limit of infinitely small steps,
leads to a diffusive dynamics with the both drift and diffusion
depending on the coordinate $x$ and time $t$.  In fact, the dynamics
arising in such case is quite similar to, for instance, overdamped
Brownian particle in a flow with the varying friction coefficient. In
this article we discussed the nontrivial dynamics arising due to this
analogy.

For instance, an exciting phenomenon arising in Brownian flows is the
stochastic ratchet effect, which allows to ``rectify'' Brownian motion
using periodically varying potential. Such potential does not
introduce any net force by itself, nevertheless allowing to push
particles in the direction opposite to the flow. As it has been shown
here, in our case we can achieve only a weak form of the stochastic
ratchet effect. That is, we can not reverse the overall drift
direction of the quantum trajectories, but only slow down this motion.
The ratchet effect manifests itself in the localization of the
''particle'' in the areas where the measurement strength is reduced
and thus the effective force is minimal.

Finally, we considered the case when the step size approaches zero as
the system approaches some state $\ket{X}$. No quantum trajectory can
cross the singularity point arising in this case.  Moreover, the
trajectories approach such singularity in the infinite time in a
similar way as they approach the ``normal'' basis states. The FP
equation demonstrate remarkable self-similarity in this case: The
arbitrary quantum walk between any subsequent zeros can be mapped to a
quantum walk between $\ket{0}$ and $\ket{1}$ with the non-vanishing
measurement strength.

The effects predicted here can be tested in the measurement-only
quantum control settings, such as, for instance, the one recently
realized experimentally using defect-in-diamond-based qubits
\cite{blok14}, but also in other setups where weak quantum measurement or
control was realized, for instance for photon-based \cite{gillett10},
ultracold-atom-based \cite{patil15,murch08} or superconducting-based
qubits \cite{murch13}.

\section*{Acknowledgment}

The author is thankful to Nieders. Vorab, project ZN3061, and to
German Research Foundation (DFG), project BA 4156/4-1, for the
financial support.
 
\appendix

 \section{\label{app:1}Derivation of the expression for step size}

 To derive the step size $\epsilon_i(x_n) \equiv x_{n+1}-x_n$ on the
 $n$th step of our random walk, we use \refeqs{eq:0}{eq:2a} and the
 relation:
 \begin{equation}
\sin^2\theta_{n} = \frac{1+\tanh x_{n}}{2}.\label{eq:sin2thetan}
\end{equation}
For instance, in the case if the measurement of the ancilla $\ket{a}$
is $\ket{0}$,  we have from \refeqs{eq:0}{eq:2a}: %\refeq{eq:4}:
\begin{equation}
 \sin{\theta_{n+1}} = \sin{\theta_n} \cos{(\delta+\alpha)}/\sqrt{p_0},
\label{eq:sinth0}
 \end{equation}
and thus, using \refeq{eq:sin2thetan}:
\begin{equation}
  \label{eq:11}
  \frac{1+\tanh x_{n+1}}{2} =\frac{(1+\tanh x_{n}) \cos^2{(\delta+\alpha) }}{2 p_0}.
\end{equation}
Hence, the expression for $\epsilon_0(x)$ (redefining $x_n$ as $x$
since $x_n$ is arbitrary) is:
\begin{equation}
  \label{eq:epsx0}
  \epsilon_0 (x) = \operatorname{atanh}{\left(\frac{(1+\tanh x) \cos^2{(\delta+\alpha)}}{p_0(x)}-1\right)} - x. 
\end{equation}
In the same way, if $\ket{a}$ collapses to 
$\ket{1}$ upon the measurement, we have
\begin{equation}
  \sin{\theta_{n+1}} = \sin{\theta_n} \sin{(\delta+\alpha)}/\sqrt{p_1},
\label{eq:sinth1}
\end{equation}
and thus we obtain for $\epsilon_1(x)$:
\begin{equation}
  \label{eq:epsx1}
  \epsilon_1 (x) = \operatorname{atanh}{\left(\frac{(1+\tanh x) \sin^2{(\delta+\alpha)}}{p_1(x)}-1\right)} - x. 
\end{equation}
Now we calculate analytically the expression for
$d\epsilon_i(x)/dx$, which can be straightforwardly shown to be zero. 
Thus, $\epsilon_i(x) =
\epsilon_i(0) \equiv \epsilon_i$ and we can take $x=0$ in
\refeq{eq:epsx0}, \refeq{eq:epsx1}, thus obtaining expressions \refeqs{eq:epsx00}{eq:epsx10}.

\section{\label{app:2} Quantum trajectories with constant
  measurement parameters}
 
In the case of a constant $x$-independent step (and only in this case) 
it is constructive to analyze $\epsilon_i$, $p_i$ on more general
level by introducing the ``average step'' $\mu(x)$:
 \begin{equation}
   \label{eq:s}
   \mu(x) = \mu_0(x)+\mu_1(x),\,\mu_i(x) = \epsilon_i p_i(x), \,i=0,1.
 \end{equation}
 $\mu$ defines an ``average direction'' of evolution: towards $+\infty$
 or $-\infty$.  

Equations \refeq{eq:s} are simplified
when $\delta$ is small (assuming fixed $\alpha>0$), so we can
decompose \refeqs{eq:p0x}{eq:epsx10} in series in $\delta$.
In this case, up to the second order of $\delta$ we have:
\begin{gather}
   \label{eq:sdeltax}
   \mu(x) =  \delta^2\tanh{(x)} + O(\delta^3),\\
   \label{eq:sdeltax0}
   2\mu_{0}(x)= \delta\sin{(2\alpha}) +
   \delta^2\left(4\Pi(x)\sin^2\alpha-1\right) + O(\delta^3), \\
   \label{eq:sdeltax1}
   2\mu_{1}(x)= -\delta \sin{(2\alpha}) +
   \delta^2\left(4\Pi(x)\cos^2\alpha-1\right) + O(\delta^3),
\end{gather}
where $\Pi(x)$ is given by \refeq{eq:pi}.  Since $\operatorname{sign}
x = \operatorname{sign} \tanh x$ \refeq{eq:sdeltax} demonstrates a
``weak attraction'' of the dynamics to the nearest state.  We also can
define in this case a quantity $D$, which have the meaning of a
diffusion coefficient:
\begin{equation}
  \label{eq:12}
  D(x) = \frac12 \sum_i p_i(x)\epsilon_i^2.  
\end{equation}
It is easy to see that in the limit of small $\delta$ (assuming
$\alpha=\mathrm{const}$) we have:
\begin{equation}
  \label{eq:13}
  D(x) = \frac12 \delta^2 +  O(\delta^3).
\end{equation}

%%%%%%%%%%%%%%%%%

 \section{\label{app:3} Derivation of the Fokker-Planck equation}

%%
 %% derivation see WBXIX:14.03.13 (note the "minus problem",
 %% WBXIX:15.03.13)
%%
%% 
%% derivation updated; prev. derivation was incorrect because the
%% steps were taken x'=x+epsilon(x), but must be x'=x+epsilon(x').
%% 

 We derive the FP equation using the standard integral
 approach \cite{gardiner09:book}. 
 Namely, we consider an arbitrary function $h(x)$ which has a finite
 support, that is, localized inside the integration area and is zero
 together with all of its derivatives for large enough $|x|$. We also
 assume that it is smooth enough. Then, we write the expression for
 $\int h(x) \partial_t P(t, x)dx$, assuming integration over the whole real axis: 
 \begin{equation}
   \label{eq:17}
   \tau_n \int h(x)  \partial_t P(t, x)dx \approxeq \int h(x) (P(t+\tau, x)-P(t, x))dx,
 \end{equation}
 Expressing $P(t+\tau_n, x)$ through $P(t, x)$ using \refeq{eq:10}
and
 assuming $t=\sum_n\tau_n$, replacing variables in two integral parts
 as $x\to x_i(x)$ followed by redefining $x_i\to x$, and finally expanding $h(x+\epsilon) \approxeq h(x) + \epsilon h'(x) +
 \epsilon^2h''(x)/2$, we transform the later expression into:
\begin{equation}
  \label{eq:18}
  \int P(t, x)\left( \sum_i p_i(h'(x)\epsilon_i(x) + h''(x)\epsilon^2_i(x)/2)\right)dx.
\end{equation}
Applying integration by parts we have finally:
\begin{gather}
\nonumber
  \int h(x) \left\{-\partial_t P(t, x) - \partial_x\left[\mu(x,t)
      P(t,x)\right] + \right.\\
 + \left.  \partial_{xx}\left[D(x,t) P(t,x)\right] \right\} dx = 0,
  \label{eq:19}
\end{gather}
where 
\begin{gather}
  \label{eq:20}
  \mu(x,t)  =
  \left.\sum_i\frac{p_i(x,n)\epsilon_i(x,n)}{\tau_n}\right|_{n\to
    t},\\
\label{eq:20a} 
D(x,t)=
\left.\sum_i\frac{p_i(x,n)\epsilon^2_i(x,n)}{2\tau_n} \right|_{n\to t}. 
\end{gather}
For small $\delta$ and $\alpha_n$ being constant
for every $n$ we have, up to the second order of $\delta$:
\begin{gather}
  \label{eq:mu:nonconst}
\sum_ip_i(x,t)\epsilon_i(x,t) = \delta_n(x)^2\tanh{(x)} + O(\delta_n(x)^3),  \\
  \label{eq:D:nonconst}
\sum_ip_i(x,t)\epsilon^2_i(x,t) = \delta_n(x)^2 +  O(\delta_n(x)^3).
\end{gather}
Taking into account \refeq{eq:delta}, we finally arrive to \refeqs{eq:25}{eq:mu:lim}.

 \section{\label{app:4} FP coefficients $\mu$ and $D$ in different coordinates}

 We may easily change the variables $x\to y(x)$ in the FP equation by the known
 rule \cite{risken84:book}:
 \begin{gather}
   \mu(y)=\mu(x) \partial_xy(x) +D(x)\partial_{xx}y(x)\label{eq:dmu}, \\
   D(y) = D(x) (\partial_xy(x))^2
   \label{eq:dy} 
 \end{gather}
 In $\theta$-coordinates, given by \refeq{eq:1}, we have:
\begin{equation}
  \label{eq:16}
\mu(\theta) = -\frac{g^2(\theta)\sin{(4\theta)} }{8},\, D(\theta) = \frac{g^2(\theta)\sin^2{(2\theta)}}{8}.  
\end{equation}
For the coordinates $\Pi(x)$ we obtain:
\begin{equation}
  \label{eq:16a}
\mu(\Pi) = 0,\, D(\Pi) = 2(\Pi-1)^2\Pi^2g^2(\Pi).
\end{equation}
The last equation for $\mu(\Pi)$ reflects the conservation of
$\langle\Pi\rangle$ as given by \refeq{eq:uppi_av}.

%%%%%%%%%

%% one should cat to wm quantum, optics, mathematics, dynamics
%\bibliographystyle{plain}
%\bibliography{wm}

\begin{thebibliography}{49}%
\makeatletter
\providecommand \@ifxundefined [1]{%
 \@ifx{#1\undefined}
}%
\providecommand \@ifnum [1]{%
 \ifnum #1\expandafter \@firstoftwo
 \else \expandafter \@secondoftwo
 \fi
}%
\providecommand \@ifx [1]{%
 \ifx #1\expandafter \@firstoftwo
 \else \expandafter \@secondoftwo
 \fi
}%
\providecommand \natexlab [1]{#1}%
\providecommand \enquote  [1]{``#1''}%
\providecommand \bibnamefont  [1]{#1}%
\providecommand \bibfnamefont [1]{#1}%
\providecommand \citenamefont [1]{#1}%
\providecommand \href@noop [0]{\@secondoftwo}%
\providecommand \href [0]{\begingroup \@sanitize@url \@href}%
\providecommand \@href[1]{\@@startlink{#1}\@@href}%
\providecommand \@@href[1]{\endgroup#1\@@endlink}%
\providecommand \@sanitize@url [0]{\catcode `\\12\catcode `\$12\catcode
  `\&12\catcode `\#12\catcode `\^12\catcode `\_12\catcode `\%12\relax}%
\providecommand \@@startlink[1]{}%
\providecommand \@@endlink[0]{}%
\providecommand \url  [0]{\begingroup\@sanitize@url \@url }%
\providecommand \@url [1]{\endgroup\@href {#1}{\urlprefix }}%
\providecommand \urlprefix  [0]{URL }%
\providecommand \Eprint [0]{\href }%
\providecommand \doibase [0]{http://dx.doi.org/}%
\providecommand \selectlanguage [0]{\@gobble}%
\providecommand \bibinfo  [0]{\@secondoftwo}%
\providecommand \bibfield  [0]{\@secondoftwo}%
\providecommand \translation [1]{[#1]}%
\providecommand \BibitemOpen [0]{}%
\providecommand \bibitemStop [0]{}%
\providecommand \bibitemNoStop [0]{.\EOS\space}%
\providecommand \EOS [0]{\spacefactor3000\relax}%
\providecommand \BibitemShut  [1]{\csname bibitem#1\endcsname}%
\let\auto@bib@innerbib\@empty
%</preamble>
\bibitem [{\citenamefont {Patil}\ \emph {et~al.}(2015)\citenamefont {Patil},
  \citenamefont {Chakram},\ and\ \citenamefont {Vengalattore}}]{patil15}%
  \BibitemOpen
  \bibfield  {author} {\bibinfo {author} {\bibfnamefont {Y.~S.}\ \bibnamefont
  {Patil}}, \bibinfo {author} {\bibfnamefont {S.}~\bibnamefont {Chakram}}, \
  and\ \bibinfo {author} {\bibfnamefont {M.}~\bibnamefont {Vengalattore}},\
  }\bibfield  {title} {\enquote {\bibinfo {title} {Measurement-induced
  localization of an ultracold lattice gas},}\ }\href {\doibase
  10.1103/PhysRevLett.115.140402} {\bibfield  {journal} {\bibinfo  {journal}
  {Phys. Rev. Lett.}\ }\textbf {\bibinfo {volume} {115}},\ \bibinfo {pages}
  {140402} (\bibinfo {year} {2015})}\BibitemShut {NoStop}%
\bibitem [{\citenamefont {Blok}\ \emph {et~al.}(2014)\citenamefont {Blok},
  \citenamefont {Bonato}, \citenamefont {Markham}, \citenamefont {Twitchen},
  \citenamefont {Dobrovitski},\ and\ \citenamefont {Hanson}}]{blok14}%
  \BibitemOpen
  \bibfield  {author} {\bibinfo {author} {\bibfnamefont {M.~S.}\ \bibnamefont
  {Blok}}, \bibinfo {author} {\bibfnamefont {C.}~\bibnamefont {Bonato}},
  \bibinfo {author} {\bibfnamefont {M.~L.}\ \bibnamefont {Markham}}, \bibinfo
  {author} {\bibfnamefont {D.~J.}\ \bibnamefont {Twitchen}}, \bibinfo {author}
  {\bibfnamefont {V.~V.}\ \bibnamefont {Dobrovitski}}, \ and\ \bibinfo {author}
  {\bibfnamefont {R.}~\bibnamefont {Hanson}},\ }\bibfield  {title} {\enquote
  {\bibinfo {title} {Manipulating a qubit through the backaction of sequential
  partial measurements and real-time feedback},}\ }\href
  {http://dx.doi.org/10.1038/nphys2881} {\bibfield  {journal} {\bibinfo
  {journal} {Nat Phys}\ }\textbf {\bibinfo {volume} {10}},\ \bibinfo {pages}
  {189--193} (\bibinfo {year} {2014})}\BibitemShut {NoStop}%
\bibitem [{\citenamefont {Murch}\ \emph {et~al.}(2013)\citenamefont {Murch},
  \citenamefont {Weber}, \citenamefont {Macklin},\ and\ \citenamefont
  {Siddiqi}}]{murch13}%
  \BibitemOpen
  \bibfield  {author} {\bibinfo {author} {\bibfnamefont {K.~W.}\ \bibnamefont
  {Murch}}, \bibinfo {author} {\bibfnamefont {S.~J.}\ \bibnamefont {Weber}},
  \bibinfo {author} {\bibfnamefont {C.}~\bibnamefont {Macklin}}, \ and\
  \bibinfo {author} {\bibfnamefont {I.}~\bibnamefont {Siddiqi}},\ }\bibfield
  {title} {\enquote {\bibinfo {title} {Observing single quantum trajectories of
  a superconducting quantum bit},}\ }\href
  {http://dx.doi.org/10.1038/nature12539} {\bibfield  {journal} {\bibinfo
  {journal} {Nature}\ }\textbf {\bibinfo {volume} {502}},\ \bibinfo {pages}
  {211--214} (\bibinfo {year} {2013})}\BibitemShut {NoStop}%
\bibitem [{\citenamefont {Hatridge}\ \emph {et~al.}(2013)\citenamefont
  {Hatridge}, \citenamefont {Shankar}, \citenamefont {Mirrahimi}, \citenamefont
  {Schackert}, \citenamefont {Geerlings}, \citenamefont {Brecht}, \citenamefont
  {Sliwa}, \citenamefont {Abdo}, \citenamefont {Frunzio}, \citenamefont
  {Girvin}, \citenamefont {Schoelkopf},\ and\ \citenamefont
  {Devoret}}]{hatridge13}%
  \BibitemOpen
  \bibfield  {author} {\bibinfo {author} {\bibfnamefont {M.}~\bibnamefont
  {Hatridge}}, \bibinfo {author} {\bibfnamefont {S.}~\bibnamefont {Shankar}},
  \bibinfo {author} {\bibfnamefont {M.}~\bibnamefont {Mirrahimi}}, \bibinfo
  {author} {\bibfnamefont {F.}~\bibnamefont {Schackert}}, \bibinfo {author}
  {\bibfnamefont {K.}~\bibnamefont {Geerlings}}, \bibinfo {author}
  {\bibfnamefont {T.}~\bibnamefont {Brecht}}, \bibinfo {author} {\bibfnamefont
  {K.~M.}\ \bibnamefont {Sliwa}}, \bibinfo {author} {\bibfnamefont
  {B.}~\bibnamefont {Abdo}}, \bibinfo {author} {\bibfnamefont {L.}~\bibnamefont
  {Frunzio}}, \bibinfo {author} {\bibfnamefont {S.~M.}\ \bibnamefont {Girvin}},
  \bibinfo {author} {\bibfnamefont {R.~J.}\ \bibnamefont {Schoelkopf}}, \ and\
  \bibinfo {author} {\bibfnamefont {M.~H.}\ \bibnamefont {Devoret}},\
  }\bibfield  {title} {\enquote {\bibinfo {title} {Quantum back-action of an
  individual variable-strength measurement},}\ }\href {\doibase
  10.1126/science.1226897} {\bibfield  {journal} {\bibinfo  {journal}
  {Science}\ }\textbf {\bibinfo {volume} {339}},\ \bibinfo {pages} {178--181}
  (\bibinfo {year} {2013})}\BibitemShut {NoStop}%
\bibitem [{\citenamefont {Wiseman}(2011)}]{wiseman11}%
  \BibitemOpen
  \bibfield  {author} {\bibinfo {author} {\bibfnamefont {Howard~M.}\
  \bibnamefont {Wiseman}},\ }\bibfield  {title} {\enquote {\bibinfo {title}
  {Quantum control: Squinting at quantum systems},}\ }\href {\doibase
  10.1038/470178a} {\bibfield  {journal} {\bibinfo  {journal} {Nature}\
  }\textbf {\bibinfo {volume} {470}},\ \bibinfo {pages} {178--179} (\bibinfo
  {year} {2011})}\BibitemShut {NoStop}%
\bibitem [{\citenamefont {Ashhab}\ and\ \citenamefont {Nori}(2010)}]{ashhab10}%
  \BibitemOpen
  \bibfield  {author} {\bibinfo {author} {\bibfnamefont {S.}~\bibnamefont
  {Ashhab}}\ and\ \bibinfo {author} {\bibfnamefont {Franco}\ \bibnamefont
  {Nori}},\ }\bibfield  {title} {\enquote {\bibinfo {title} {Control-free
  control: Manipulating a quantum system using only a limited set of
  measurements},}\ }\href {\doibase 10.1103/PhysRevA.82.062103} {\bibfield
  {journal} {\bibinfo  {journal} {Phys. Rev. A}\ }\textbf {\bibinfo {volume}
  {82}},\ \bibinfo {pages} {062103} (\bibinfo {year} {2010})}\BibitemShut
  {NoStop}%
\bibitem [{\citenamefont {Guerlin}\ \emph {et~al.}(2007)\citenamefont
  {Guerlin}, \citenamefont {Bernu}, \citenamefont {Del{\'e}glise},
  \citenamefont {Sayrin}, \citenamefont {Gleyzes}, \citenamefont {Kuhr},
  \citenamefont {Brune}, \citenamefont {Raimond},\ and\ \citenamefont
  {Haroche}}]{guerlin07}%
  \BibitemOpen
  \bibfield  {author} {\bibinfo {author} {\bibfnamefont {Christine}\
  \bibnamefont {Guerlin}}, \bibinfo {author} {\bibfnamefont {Julien}\
  \bibnamefont {Bernu}}, \bibinfo {author} {\bibfnamefont {Samuel}\
  \bibnamefont {Del{\'e}glise}}, \bibinfo {author} {\bibfnamefont
  {Cl{\'e}ment}\ \bibnamefont {Sayrin}}, \bibinfo {author} {\bibfnamefont
  {S{\'e}bastien}\ \bibnamefont {Gleyzes}}, \bibinfo {author} {\bibfnamefont
  {Stefan}\ \bibnamefont {Kuhr}}, \bibinfo {author} {\bibfnamefont {Michel}\
  \bibnamefont {Brune}}, \bibinfo {author} {\bibfnamefont {Jean-Michel}\
  \bibnamefont {Raimond}}, \ and\ \bibinfo {author} {\bibfnamefont {Serge}\
  \bibnamefont {Haroche}},\ }\bibfield  {title} {\enquote {\bibinfo {title}
  {Progressive field-state collapse and quantum non-demolition photon
  counting},}\ }\href@noop {} {\bibfield  {journal} {\bibinfo  {journal}
  {Nature}\ }\textbf {\bibinfo {volume} {448}},\ \bibinfo {pages} {889--893}
  (\bibinfo {year} {2007})}\BibitemShut {NoStop}%
\bibitem [{\citenamefont {Gleyzes}\ \emph {et~al.}(2007)\citenamefont
  {Gleyzes}, \citenamefont {Kuhr}, \citenamefont {Guerlin}, \citenamefont
  {Bernu}, \citenamefont {Deleglise}, \citenamefont {Busk~Hoff}, \citenamefont
  {Brune}, \citenamefont {Raimond},\ and\ \citenamefont {Haroche}}]{gleyzes07}%
  \BibitemOpen
  \bibfield  {author} {\bibinfo {author} {\bibfnamefont {Sebastien}\
  \bibnamefont {Gleyzes}}, \bibinfo {author} {\bibfnamefont {Stefan}\
  \bibnamefont {Kuhr}}, \bibinfo {author} {\bibfnamefont {Christine}\
  \bibnamefont {Guerlin}}, \bibinfo {author} {\bibfnamefont {Julien}\
  \bibnamefont {Bernu}}, \bibinfo {author} {\bibfnamefont {Samuel}\
  \bibnamefont {Deleglise}}, \bibinfo {author} {\bibfnamefont {Ulrich}\
  \bibnamefont {Busk~Hoff}}, \bibinfo {author} {\bibfnamefont {Michel}\
  \bibnamefont {Brune}}, \bibinfo {author} {\bibfnamefont {Jean-Michel}\
  \bibnamefont {Raimond}}, \ and\ \bibinfo {author} {\bibfnamefont {Serge}\
  \bibnamefont {Haroche}},\ }\bibfield  {title} {\enquote {\bibinfo {title}
  {Quantum jumps of light recording the birth and death of a photon in a
  cavity},}\ }\href {\doibase 10.1038/nature05589} {\bibfield  {journal}
  {\bibinfo  {journal} {Nature}\ }\textbf {\bibinfo {volume} {446}},\ \bibinfo
  {pages} {297--300} (\bibinfo {year} {2007})}\BibitemShut {NoStop}%
\bibitem [{\citenamefont {Murch}\ \emph {et~al.}(2008)\citenamefont {Murch},
  \citenamefont {Moore}, \citenamefont {Gupta},\ and\ \citenamefont
  {Stamper-Kurn}}]{murch08}%
  \BibitemOpen
  \bibfield  {author} {\bibinfo {author} {\bibfnamefont {Kater~W.}\
  \bibnamefont {Murch}}, \bibinfo {author} {\bibfnamefont {Kevin~L.}\
  \bibnamefont {Moore}}, \bibinfo {author} {\bibfnamefont {Subhadeep}\
  \bibnamefont {Gupta}}, \ and\ \bibinfo {author} {\bibfnamefont {Dan~M.}\
  \bibnamefont {Stamper-Kurn}},\ }\bibfield  {title} {\enquote {\bibinfo
  {title} {Observation of quantum-measurement backaction with an ultracold
  atomic gas},}\ }\href {\doibase 10.1038/nphys965} {\bibfield  {journal}
  {\bibinfo  {journal} {Nat Phys}\ }\textbf {\bibinfo {volume} {4}},\ \bibinfo
  {pages} {561--564} (\bibinfo {year} {2008})}\BibitemShut {NoStop}%
\bibitem [{\citenamefont {Gammelmark}\ \emph {et~al.}(2013)\citenamefont
  {Gammelmark}, \citenamefont {Julsgaard},\ and\ \citenamefont
  {M\o{}lmer}}]{gammelmark13}%
  \BibitemOpen
  \bibfield  {author} {\bibinfo {author} {\bibfnamefont {S\o{}ren}\
  \bibnamefont {Gammelmark}}, \bibinfo {author} {\bibfnamefont {Brian}\
  \bibnamefont {Julsgaard}}, \ and\ \bibinfo {author} {\bibfnamefont {Klaus}\
  \bibnamefont {M\o{}lmer}},\ }\bibfield  {title} {\enquote {\bibinfo {title}
  {Past quantum states of a monitored system},}\ }\href {\doibase
  10.1103/PhysRevLett.111.160401} {\bibfield  {journal} {\bibinfo  {journal}
  {Phys. Rev. Lett.}\ }\textbf {\bibinfo {volume} {111}},\ \bibinfo {pages}
  {160401} (\bibinfo {year} {2013})}\BibitemShut {NoStop}%
\bibitem [{\citenamefont {Pechen}\ \emph {et~al.}(2006)\citenamefont {Pechen},
  \citenamefont {Il'in}, \citenamefont {Shuang},\ and\ \citenamefont
  {Rabitz}}]{pechen06}%
  \BibitemOpen
  \bibfield  {author} {\bibinfo {author} {\bibfnamefont {Alexander}\
  \bibnamefont {Pechen}}, \bibinfo {author} {\bibfnamefont {Nikolai}\
  \bibnamefont {Il'in}}, \bibinfo {author} {\bibfnamefont {Feng}\ \bibnamefont
  {Shuang}}, \ and\ \bibinfo {author} {\bibfnamefont {Herschel}\ \bibnamefont
  {Rabitz}},\ }\bibfield  {title} {\enquote {\bibinfo {title} {Quantum control
  by von neumann measurements},}\ }\href {\doibase 10.1103/PhysRevA.74.052102}
  {\bibfield  {journal} {\bibinfo  {journal} {Phys. Rev. A}\ }\textbf {\bibinfo
  {volume} {74}},\ \bibinfo {pages} {052102} (\bibinfo {year}
  {2006})}\BibitemShut {NoStop}%
\bibitem [{\citenamefont {Gordon}\ \emph {et~al.}(2013)\citenamefont {Gordon},
  \citenamefont {Mazets},\ and\ \citenamefont {Kurizki}}]{gordon13}%
  \BibitemOpen
  \bibfield  {author} {\bibinfo {author} {\bibfnamefont {Goren}\ \bibnamefont
  {Gordon}}, \bibinfo {author} {\bibfnamefont {Igor~E.}\ \bibnamefont
  {Mazets}}, \ and\ \bibinfo {author} {\bibfnamefont {Gershon}\ \bibnamefont
  {Kurizki}},\ }\bibfield  {title} {\enquote {\bibinfo {title} {Quantum
  particle localization by frequent coherent monitoring},}\ }\href {\doibase
  10.1103/PhysRevA.87.052141} {\bibfield  {journal} {\bibinfo  {journal} {Phys.
  Rev. A}\ }\textbf {\bibinfo {volume} {87}},\ \bibinfo {pages} {052141}
  (\bibinfo {year} {2013})}\BibitemShut {NoStop}%
\bibitem [{\citenamefont {Mackrory}\ \emph {et~al.}(2010)\citenamefont
  {Mackrory}, \citenamefont {Jacobs},\ and\ \citenamefont
  {Steck}}]{mackrory10}%
  \BibitemOpen
  \bibfield  {author} {\bibinfo {author} {\bibfnamefont {Jonathan~B}\
  \bibnamefont {Mackrory}}, \bibinfo {author} {\bibfnamefont {Kurt}\
  \bibnamefont {Jacobs}}, \ and\ \bibinfo {author} {\bibfnamefont {Daniel~A}\
  \bibnamefont {Steck}},\ }\bibfield  {title} {\enquote {\bibinfo {title}
  {Reflection of a particle from a quantum measurement},}\ }\href
  {http://stacks.iop.org/1367-2630/12/i=11/a=113023} {\bibfield  {journal}
  {\bibinfo  {journal} {New Journal of Physics}\ }\textbf {\bibinfo {volume}
  {12}},\ \bibinfo {pages} {113023} (\bibinfo {year} {2010})}\BibitemShut
  {NoStop}%
\bibitem [{\citenamefont {Wiseman}\ and\ \citenamefont
  {Milburn}(2010)}]{wiseman10:book}%
  \BibitemOpen
  \bibfield  {author} {\bibinfo {author} {\bibfnamefont {H.M.}\ \bibnamefont
  {Wiseman}}\ and\ \bibinfo {author} {\bibfnamefont {G.J.}\ \bibnamefont
  {Milburn}},\ }\href {http://books.google.de/books?id=ZNjvHaH8qA4C} {\emph
  {\bibinfo {title} {Quantum Measurement and Control}}}\ (\bibinfo  {publisher}
  {Cambridge University Press},\ \bibinfo {year} {2010})\BibitemShut {NoStop}%
\bibitem [{\citenamefont {Paz-Silva}\ \emph {et~al.}(2012)\citenamefont
  {Paz-Silva}, \citenamefont {Rezakhani}, \citenamefont {Dominy},\ and\
  \citenamefont {Lidar}}]{paz-silva12}%
  \BibitemOpen
  \bibfield  {author} {\bibinfo {author} {\bibfnamefont {Gerardo~A.}\
  \bibnamefont {Paz-Silva}}, \bibinfo {author} {\bibfnamefont {A.~T.}\
  \bibnamefont {Rezakhani}}, \bibinfo {author} {\bibfnamefont {Jason~M.}\
  \bibnamefont {Dominy}}, \ and\ \bibinfo {author} {\bibfnamefont {D.~A.}\
  \bibnamefont {Lidar}},\ }\bibfield  {title} {\enquote {\bibinfo {title} {Zeno
  effect for quantum computation and control},}\ }\href {\doibase
  10.1103/PhysRevLett.108.080501} {\bibfield  {journal} {\bibinfo  {journal}
  {Phys. Rev. Lett.}\ }\textbf {\bibinfo {volume} {108}},\ \bibinfo {pages}
  {080501} (\bibinfo {year} {2012})}\BibitemShut {NoStop}%
\bibitem [{\citenamefont {Caves}\ and\ \citenamefont
  {Milburn}(1987)}]{caves87}%
  \BibitemOpen
  \bibfield  {author} {\bibinfo {author} {\bibfnamefont {Carlton~M.}\
  \bibnamefont {Caves}}\ and\ \bibinfo {author} {\bibfnamefont {G.~J.}\
  \bibnamefont {Milburn}},\ }\bibfield  {title} {\enquote {\bibinfo {title}
  {Quantum-mechanical model for continuous position measurements},}\ }\href
  {\doibase 10.1103/PhysRevA.36.5543} {\bibfield  {journal} {\bibinfo
  {journal} {Phys. Rev. A}\ }\textbf {\bibinfo {volume} {36}},\ \bibinfo
  {pages} {5543--5555} (\bibinfo {year} {1987})}\BibitemShut {NoStop}%
\bibitem [{\citenamefont {Oreshkov}\ and\ \citenamefont
  {Brun}(2005)}]{oreshkov05}%
  \BibitemOpen
  \bibfield  {author} {\bibinfo {author} {\bibfnamefont {Ognyan}\ \bibnamefont
  {Oreshkov}}\ and\ \bibinfo {author} {\bibfnamefont {Todd~A.}\ \bibnamefont
  {Brun}},\ }\bibfield  {title} {\enquote {\bibinfo {title} {Weak measurements
  are universal},}\ }\href {\doibase 10.1103/PhysRevLett.95.110409} {\bibfield
  {journal} {\bibinfo  {journal} {Phys. Rev. Lett.}\ }\textbf {\bibinfo
  {volume} {95}},\ \bibinfo {pages} {110409} (\bibinfo {year}
  {2005})}\BibitemShut {NoStop}%
\bibitem [{\citenamefont {Varbanov}\ and\ \citenamefont
  {Brun}(2007)}]{varbanov07}%
  \BibitemOpen
  \bibfield  {author} {\bibinfo {author} {\bibfnamefont {Martin}\ \bibnamefont
  {Varbanov}}\ and\ \bibinfo {author} {\bibfnamefont {Todd~A.}\ \bibnamefont
  {Brun}},\ }\bibfield  {title} {\enquote {\bibinfo {title} {Decomposing
  generalized measurements into continuous stochastic processes},}\ }\href
  {\doibase 10.1103/PhysRevA.76.032104} {\bibfield  {journal} {\bibinfo
  {journal} {Phys. Rev. A}\ }\textbf {\bibinfo {volume} {76}},\ \bibinfo
  {pages} {032104} (\bibinfo {year} {2007})}\BibitemShut {NoStop}%
\bibitem [{\citenamefont {Aharonov}\ \emph {et~al.}(1988)\citenamefont
  {Aharonov}, \citenamefont {Albert},\ and\ \citenamefont
  {Vaidman}}]{aharonov88}%
  \BibitemOpen
  \bibfield  {author} {\bibinfo {author} {\bibfnamefont {Yakir}\ \bibnamefont
  {Aharonov}}, \bibinfo {author} {\bibfnamefont {David~Z.}\ \bibnamefont
  {Albert}}, \ and\ \bibinfo {author} {\bibfnamefont {Lev}\ \bibnamefont
  {Vaidman}},\ }\bibfield  {title} {\enquote {\bibinfo {title} {How the result
  of a measurement of a component of the spin of a spin-1/2 particle can turn
  out to be 100},}\ }\href {\doibase 10.1103/PhysRevLett.60.1351} {\bibfield
  {journal} {\bibinfo  {journal} {Phys. Rev. Lett.}\ }\textbf {\bibinfo
  {volume} {60}},\ \bibinfo {pages} {1351--1354} (\bibinfo {year}
  {1988})}\BibitemShut {NoStop}%
\bibitem [{\citenamefont {Dressel}\ and\ \citenamefont
  {Jordan}(2012)}]{dressel12}%
  \BibitemOpen
  \bibfield  {author} {\bibinfo {author} {\bibfnamefont {Justin}\ \bibnamefont
  {Dressel}}\ and\ \bibinfo {author} {\bibfnamefont {Andrew~N.}\ \bibnamefont
  {Jordan}},\ }\bibfield  {title} {\enquote {\bibinfo {title} {Weak values are
  universal in von neumann measurements},}\ }\href {\doibase
  10.1103/PhysRevLett.109.230402} {\bibfield  {journal} {\bibinfo  {journal}
  {Phys. Rev. Lett.}\ }\textbf {\bibinfo {volume} {109}},\ \bibinfo {pages}
  {230402} (\bibinfo {year} {2012})}\BibitemShut {NoStop}%
\bibitem [{\citenamefont {Lundeen}\ and\ \citenamefont
  {Bamber}(2012)}]{lundeen12}%
  \BibitemOpen
  \bibfield  {author} {\bibinfo {author} {\bibfnamefont {Jeff~S.}\ \bibnamefont
  {Lundeen}}\ and\ \bibinfo {author} {\bibfnamefont {Charles}\ \bibnamefont
  {Bamber}},\ }\bibfield  {title} {\enquote {\bibinfo {title} {Procedure for
  direct measurement of general quantum states using weak measurement},}\
  }\href {\doibase 10.1103/PhysRevLett.108.070402} {\bibfield  {journal}
  {\bibinfo  {journal} {Phys. Rev. Lett.}\ }\textbf {\bibinfo {volume} {108}},\
  \bibinfo {pages} {070402} (\bibinfo {year} {2012})}\BibitemShut {NoStop}%
\bibitem [{\citenamefont {Belavkin}(1989)}]{belavkin89}%
  \BibitemOpen
  \bibfield  {author} {\bibinfo {author} {\bibfnamefont {V.P.}\ \bibnamefont
  {Belavkin}},\ }\bibfield  {title} {\enquote {\bibinfo {title} {A new wave
  equation for a continuous nondemolition measurement},}\ }\href {\doibase
  10.1016/0375-9601(89)90066-2} {\bibfield  {journal} {\bibinfo  {journal}
  {Physics Letters A}\ }\textbf {\bibinfo {volume} {140}},\ \bibinfo {pages}
  {355 -- 358} (\bibinfo {year} {1989})}\BibitemShut {NoStop}%
\bibitem [{\citenamefont {Gisin}(1984)}]{gisin84}%
  \BibitemOpen
  \bibfield  {author} {\bibinfo {author} {\bibfnamefont {N.}~\bibnamefont
  {Gisin}},\ }\bibfield  {title} {\enquote {\bibinfo {title} {Quantum
  measurements and stochastic processes},}\ }\href {\doibase
  10.1103/PhysRevLett.52.1657} {\bibfield  {journal} {\bibinfo  {journal}
  {Phys. Rev. Lett.}\ }\textbf {\bibinfo {volume} {52}},\ \bibinfo {pages}
  {1657--1660} (\bibinfo {year} {1984})}\BibitemShut {NoStop}%
\bibitem [{\citenamefont {Wiseman}(1996)}]{wiseman96}%
  \BibitemOpen
  \bibfield  {author} {\bibinfo {author} {\bibfnamefont {H~M}\ \bibnamefont
  {Wiseman}},\ }\bibfield  {title} {\enquote {\bibinfo {title} {Quantum
  trajectories and quantum measurement theory},}\ }\href
  {http://stacks.iop.org/1355-5111/8/i=1/a=015} {\bibfield  {journal} {\bibinfo
   {journal} {Quantum and Semiclassical Optics: Journal of the European Optical
  Society Part B}\ }\textbf {\bibinfo {volume} {8}},\ \bibinfo {pages} {205}
  (\bibinfo {year} {1996})}\BibitemShut {NoStop}%
\bibitem [{\citenamefont {Gillett}\ \emph {et~al.}(2010)\citenamefont
  {Gillett}, \citenamefont {Dalton}, \citenamefont {Lanyon}, \citenamefont
  {Almeida}, \citenamefont {Barbieri}, \citenamefont {Pryde}, \citenamefont
  {O'Brien}, \citenamefont {Resch}, \citenamefont {Bartlett},\ and\
  \citenamefont {White}}]{gillett10}%
  \BibitemOpen
  \bibfield  {author} {\bibinfo {author} {\bibfnamefont {G.~G.}\ \bibnamefont
  {Gillett}}, \bibinfo {author} {\bibfnamefont {R.~B.}\ \bibnamefont {Dalton}},
  \bibinfo {author} {\bibfnamefont {B.~P.}\ \bibnamefont {Lanyon}}, \bibinfo
  {author} {\bibfnamefont {M.~P.}\ \bibnamefont {Almeida}}, \bibinfo {author}
  {\bibfnamefont {M.}~\bibnamefont {Barbieri}}, \bibinfo {author}
  {\bibfnamefont {G.~J.}\ \bibnamefont {Pryde}}, \bibinfo {author}
  {\bibfnamefont {J.~L.}\ \bibnamefont {O'Brien}}, \bibinfo {author}
  {\bibfnamefont {K.~J.}\ \bibnamefont {Resch}}, \bibinfo {author}
  {\bibfnamefont {S.~D.}\ \bibnamefont {Bartlett}}, \ and\ \bibinfo {author}
  {\bibfnamefont {A.~G.}\ \bibnamefont {White}},\ }\bibfield  {title} {\enquote
  {\bibinfo {title} {Experimental feedback control of quantum systems using
  weak measurements},}\ }\href {\doibase 10.1103/PhysRevLett.104.080503}
  {\bibfield  {journal} {\bibinfo  {journal} {Phys. Rev. Lett.}\ }\textbf
  {\bibinfo {volume} {104}},\ \bibinfo {pages} {080503} (\bibinfo {year}
  {2010})}\BibitemShut {NoStop}%
\bibitem [{\citenamefont {Karasik}\ and\ \citenamefont
  {Wiseman}(2011)}]{karasik11}%
  \BibitemOpen
  \bibfield  {author} {\bibinfo {author} {\bibfnamefont {R.~I.}\ \bibnamefont
  {Karasik}}\ and\ \bibinfo {author} {\bibfnamefont {H.~M.}\ \bibnamefont
  {Wiseman}},\ }\bibfield  {title} {\enquote {\bibinfo {title} {How many bits
  does it take to track an open quantum system?}}\ }\href {\doibase
  10.1103/PhysRevLett.106.020406} {\bibfield  {journal} {\bibinfo  {journal}
  {Phys. Rev. Lett.}\ }\textbf {\bibinfo {volume} {106}},\ \bibinfo {pages}
  {020406} (\bibinfo {year} {2011})}\BibitemShut {NoStop}%
\bibitem [{\citenamefont {H{\"a}nggi}\ and\ \citenamefont
  {Marchesoni}(2009)}]{haenggi09}%
  \BibitemOpen
  \bibfield  {author} {\bibinfo {author} {\bibfnamefont {Peter}\ \bibnamefont
  {H{\"a}nggi}}\ and\ \bibinfo {author} {\bibfnamefont {Fabio}\ \bibnamefont
  {Marchesoni}},\ }\bibfield  {title} {\enquote {\bibinfo {title} {Artificial
  brownian motors: Controlling transport on the nanoscale},}\ }\href@noop {}
  {\bibfield  {journal} {\bibinfo  {journal} {Reviews of Modern Physics}\
  }\textbf {\bibinfo {volume} {81}},\ \bibinfo {pages} {387} (\bibinfo {year}
  {2009})}\BibitemShut {NoStop}%
\bibitem [{\citenamefont {Reimann}(2002)}]{reimann02}%
  \BibitemOpen
  \bibfield  {author} {\bibinfo {author} {\bibfnamefont {Peter}\ \bibnamefont
  {Reimann}},\ }\bibfield  {title} {\enquote {\bibinfo {title} {Brownian
  motors: noisy transport far from equilibrium},}\ }\href {\doibase
  10.1016/S0370-1573(01)00081-3} {\bibfield  {journal} {\bibinfo  {journal}
  {Physics Reports}\ }\textbf {\bibinfo {volume} {361}},\ \bibinfo {pages} {57
  -- 265} (\bibinfo {year} {2002})}\BibitemShut {NoStop}%
\bibitem [{\citenamefont {Braun}\ and\ \citenamefont
  {Kivshar}(2004)}]{braun:book04}%
  \BibitemOpen
  \bibfield  {author} {\bibinfo {author} {\bibfnamefont {O.M.}\ \bibnamefont
  {Braun}}\ and\ \bibinfo {author} {\bibfnamefont {Y.S.}\ \bibnamefont
  {Kivshar}},\ }\href {https://books.google.de/books?id=zyoT068mu0YC} {\emph
  {\bibinfo {title} {The Frenkel-Kontorova Model: Concepts, Methods, and
  Applications}}},\ Physics and Astronomy Online Library\ (\bibinfo
  {publisher} {Springer},\ \bibinfo {year} {2004})\BibitemShut {NoStop}%
\bibitem [{\citenamefont {Wu}\ and\ \citenamefont {Szeto}(2014)}]{wu14}%
  \BibitemOpen
  \bibfield  {author} {\bibinfo {author} {\bibfnamefont {Degang}\ \bibnamefont
  {Wu}}\ and\ \bibinfo {author} {\bibfnamefont {Kwok~Yip}\ \bibnamefont
  {Szeto}},\ }\bibfield  {title} {\enquote {\bibinfo {title} {Extended
  {P}arrondo's game and {B}rownian ratchets: Strong and weak {P}arrondo
  effect},}\ }\href {\doibase 10.1103/PhysRevE.89.022142} {\bibfield  {journal}
  {\bibinfo  {journal} {Phys. Rev. E}\ }\textbf {\bibinfo {volume} {89}},\
  \bibinfo {pages} {022142} (\bibinfo {year} {2014})}\BibitemShut {NoStop}%
\bibitem [{\citenamefont {Velez}\ \emph {et~al.}(2008)\citenamefont {Velez},
  \citenamefont {Martin}, \citenamefont {Villegas}, \citenamefont {Hoffmann},
  \citenamefont {Gonz{\'a}lez}, \citenamefont {Vicent},\ and\ \citenamefont
  {Schuller}}]{velez08}%
  \BibitemOpen
  \bibfield  {author} {\bibinfo {author} {\bibfnamefont {M}~\bibnamefont
  {Velez}}, \bibinfo {author} {\bibfnamefont {JI}~\bibnamefont {Martin}},
  \bibinfo {author} {\bibfnamefont {JE}~\bibnamefont {Villegas}}, \bibinfo
  {author} {\bibfnamefont {Axel}\ \bibnamefont {Hoffmann}}, \bibinfo {author}
  {\bibfnamefont {EM}~\bibnamefont {Gonz{\'a}lez}}, \bibinfo {author}
  {\bibfnamefont {JL}~\bibnamefont {Vicent}}, \ and\ \bibinfo {author}
  {\bibfnamefont {Ivan~K}\ \bibnamefont {Schuller}},\ }\bibfield  {title}
  {\enquote {\bibinfo {title} {Superconducting vortex pinning with artificial
  magnetic nanostructures},}\ }\href@noop {} {\bibfield  {journal} {\bibinfo
  {journal} {Journal of Magnetism and Magnetic Materials}\ }\textbf {\bibinfo
  {volume} {320}},\ \bibinfo {pages} {2547--2562} (\bibinfo {year}
  {2008})}\BibitemShut {NoStop}%
\bibitem [{\citenamefont {de~Souza~Silva}\ \emph {et~al.}(2006)\citenamefont
  {de~Souza~Silva}, \citenamefont {Van~de Vondel}, \citenamefont {Morelle},\
  and\ \citenamefont {Moshchalkov}}]{souza06}%
  \BibitemOpen
  \bibfield  {author} {\bibinfo {author} {\bibfnamefont {Cl{\'e}cio~C}\
  \bibnamefont {de~Souza~Silva}}, \bibinfo {author} {\bibfnamefont {Joris}\
  \bibnamefont {Van~de Vondel}}, \bibinfo {author} {\bibfnamefont {Mathieu}\
  \bibnamefont {Morelle}}, \ and\ \bibinfo {author} {\bibfnamefont {Victor~V}\
  \bibnamefont {Moshchalkov}},\ }\bibfield  {title} {\enquote {\bibinfo {title}
  {Controlled multiple reversals of a ratchet effect},}\ }\href@noop {}
  {\bibfield  {journal} {\bibinfo  {journal} {Nature}\ }\textbf {\bibinfo
  {volume} {440}},\ \bibinfo {pages} {651--654} (\bibinfo {year}
  {2006})}\BibitemShut {NoStop}%
\bibitem [{\citenamefont {Villegas}\ \emph {et~al.}(2005)\citenamefont
  {Villegas}, \citenamefont {Gonzalez}, \citenamefont {Gonzalez}, \citenamefont
  {Anguita},\ and\ \citenamefont {Vicent}}]{villegas05}%
  \BibitemOpen
  \bibfield  {author} {\bibinfo {author} {\bibfnamefont {JE}~\bibnamefont
  {Villegas}}, \bibinfo {author} {\bibfnamefont {EM}~\bibnamefont {Gonzalez}},
  \bibinfo {author} {\bibfnamefont {MP}~\bibnamefont {Gonzalez}}, \bibinfo
  {author} {\bibfnamefont {Jos{\'e}~Virgilio}\ \bibnamefont {Anguita}}, \ and\
  \bibinfo {author} {\bibfnamefont {JL}~\bibnamefont {Vicent}},\ }\bibfield
  {title} {\enquote {\bibinfo {title} {Experimental ratchet effect in
  superconducting films with periodic arrays of asymmetric potentials},}\
  }\href@noop {} {\bibfield  {journal} {\bibinfo  {journal} {Physical Review
  B}\ }\textbf {\bibinfo {volume} {71}},\ \bibinfo {pages} {024519} (\bibinfo
  {year} {2005})}\BibitemShut {NoStop}%
\bibitem [{\citenamefont {Parrondo}(1996)}]{parrondo96}%
  \BibitemOpen
  \bibfield  {author} {\bibinfo {author} {\bibfnamefont {J.~M.~R.}\
  \bibnamefont {Parrondo}},\ }\bibfield  {title} {\enquote {\bibinfo {title}
  {How to cheat a bad mathematician},}\ }in\ \href@noop {} {\emph {\bibinfo
  {booktitle} {EEC HC{\&}M Network on Complexity and Chaos}}}\ (\bibinfo
  {address} {ISI, Torino, Italy},\ \bibinfo {year} {1996})\BibitemShut
  {NoStop}%
\bibitem [{\citenamefont {Allison}\ and\ \citenamefont
  {Abbott}(2002)}]{allison02}%
  \BibitemOpen
  \bibfield  {author} {\bibinfo {author} {\bibfnamefont {Andrew}\ \bibnamefont
  {Allison}}\ and\ \bibinfo {author} {\bibfnamefont {Derek}\ \bibnamefont
  {Abbott}},\ }\bibfield  {title} {\enquote {\bibinfo {title} {The physical
  basis for {P}arrondo's games},}\ }\href@noop {} {\bibfield  {journal}
  {\bibinfo  {journal} {Fluctuation and Noise Letters}\ }\textbf {\bibinfo
  {volume} {2}},\ \bibinfo {pages} {L327--L341} (\bibinfo {year}
  {2002})}\BibitemShut {NoStop}%
\bibitem [{\citenamefont {Pawela}\ and\ \citenamefont
  {S{\l}adkowski}(2013)}]{pawela13}%
  \BibitemOpen
  \bibfield  {author} {\bibinfo {author} {\bibfnamefont {{\L}ukasz}\
  \bibnamefont {Pawela}}\ and\ \bibinfo {author} {\bibfnamefont {Jan}\
  \bibnamefont {S{\l}adkowski}},\ }\bibfield  {title} {\enquote {\bibinfo
  {title} {Cooperative quantum {P}arrondo’s games},}\ }\href@noop {}
  {\bibfield  {journal} {\bibinfo  {journal} {Physica D: Nonlinear Phenomena}\
  }\textbf {\bibinfo {volume} {256}},\ \bibinfo {pages} {51--57} (\bibinfo
  {year} {2013})}\BibitemShut {NoStop}%
\bibitem [{\citenamefont {Chandrashekar}\ and\ \citenamefont
  {Banerjee}(2011)}]{chandrashekar11}%
  \BibitemOpen
  \bibfield  {author} {\bibinfo {author} {\bibfnamefont {C.M.}\ \bibnamefont
  {Chandrashekar}}\ and\ \bibinfo {author} {\bibfnamefont {Subhashish}\
  \bibnamefont {Banerjee}},\ }\bibfield  {title} {\enquote {\bibinfo {title}
  {Parrondoʼs game using a discrete-time quantum walk},}\ }\href {\doibase
  http://dx.doi.org/10.1016/j.physleta.2011.02.071} {\bibfield  {journal}
  {\bibinfo  {journal} {Physics Letters A}\ }\textbf {\bibinfo {volume}
  {375}},\ \bibinfo {pages} {1553 -- 1558} (\bibinfo {year}
  {2011})}\BibitemShut {NoStop}%
\bibitem [{\citenamefont {Bulger}\ \emph {et~al.}(2008)\citenamefont {Bulger},
  \citenamefont {Freckleton},\ and\ \citenamefont {Twamley}}]{bulger08}%
  \BibitemOpen
  \bibfield  {author} {\bibinfo {author} {\bibfnamefont {David}\ \bibnamefont
  {Bulger}}, \bibinfo {author} {\bibfnamefont {James}\ \bibnamefont
  {Freckleton}}, \ and\ \bibinfo {author} {\bibfnamefont {Jason}\ \bibnamefont
  {Twamley}},\ }\bibfield  {title} {\enquote {\bibinfo {title}
  {Position-dependent and cooperative quantum {P}arrondo walks},}\ }\href
  {http://stacks.iop.org/1367-2630/10/i=9/a=093014} {\bibfield  {journal}
  {\bibinfo  {journal} {New Journal of Physics}\ }\textbf {\bibinfo {volume}
  {10}},\ \bibinfo {pages} {093014} (\bibinfo {year} {2008})}\BibitemShut
  {NoStop}%
\bibitem [{\citenamefont {Flitney}\ \emph {et~al.}(2002)\citenamefont
  {Flitney}, \citenamefont {Ng},\ and\ \citenamefont {Abbott}}]{flitney02}%
  \BibitemOpen
  \bibfield  {author} {\bibinfo {author} {\bibfnamefont {A.P.}\ \bibnamefont
  {Flitney}}, \bibinfo {author} {\bibfnamefont {J.}~\bibnamefont {Ng}}, \ and\
  \bibinfo {author} {\bibfnamefont {D.}~\bibnamefont {Abbott}},\ }\bibfield
  {title} {\enquote {\bibinfo {title} {Quantum {P}arrondo's games},}\ }\href
  {\doibase http://dx.doi.org/10.1016/S0378-4371(02)01084-1} {\bibfield
  {journal} {\bibinfo  {journal} {Physica A: Statistical Mechanics and its
  Applications}\ }\textbf {\bibinfo {volume} {314}},\ \bibinfo {pages} {35 --
  42} (\bibinfo {year} {2002})}\BibitemShut {NoStop}%
\bibitem [{\citenamefont {Gardiner}\ and\ \citenamefont
  {Gardiner}(2009)}]{gardiner09:book}%
  \BibitemOpen
  \bibfield  {author} {\bibinfo {author} {\bibfnamefont {Crispin~W}\
  \bibnamefont {Gardiner}}\ and\ \bibinfo {author} {\bibfnamefont
  {C}~\bibnamefont {Gardiner}},\ }\href@noop {} {\emph {\bibinfo {title}
  {Stochastic methods: a handbook for the natural and social sciences}}},\
  Vol.~\bibinfo {volume} {4}\ (\bibinfo  {publisher} {Springer Berlin},\
  \bibinfo {year} {2009})\BibitemShut {NoStop}%
\bibitem [{\citenamefont {{Brun}}(2002)}]{brun02a}%
  \BibitemOpen
  \bibfield  {author} {\bibinfo {author} {\bibfnamefont {T.~A.}\ \bibnamefont
  {{Brun}}},\ }\bibfield  {title} {\enquote {\bibinfo {title} {{A simple model
  of quantum trajectories}},}\ }\href {\doibase 10.1119/1.1475328} {\bibfield
  {journal} {\bibinfo  {journal} {American Journal of Physics}\ }\textbf
  {\bibinfo {volume} {70}},\ \bibinfo {pages} {719--737} (\bibinfo {year}
  {2002})},\ \Eprint {http://arxiv.org/abs/arXiv:quant-ph/0108132}
  {arXiv:quant-ph/0108132} \BibitemShut {NoStop}%
\bibitem [{\citenamefont {Bauer}\ \emph {et~al.}(2013)\citenamefont {Bauer},
  \citenamefont {Benoist},\ and\ \citenamefont {Bernard}}]{bauer13}%
  \BibitemOpen
  \bibfield  {author} {\bibinfo {author} {\bibfnamefont {Michel}\ \bibnamefont
  {Bauer}}, \bibinfo {author} {\bibfnamefont {Tristan}\ \bibnamefont
  {Benoist}}, \ and\ \bibinfo {author} {\bibfnamefont {Denis}\ \bibnamefont
  {Bernard}},\ }\bibfield  {title} {\enquote {\bibinfo {title} {Repeated
  quantum non-demolition measurements: Convergence and continuous time
  limit},}\ }in\ \href@noop {} {\emph {\bibinfo {booktitle} {Annales Henri
  Poincar{\'e}}}}\ (\bibinfo {organization} {Springer},\ \bibinfo {year}
  {2013})\ pp.\ \bibinfo {pages} {1--41}\BibitemShut {NoStop}%
\bibitem [{\citenamefont {Doering}(1998)}]{doering98}%
  \BibitemOpen
  \bibfield  {author} {\bibinfo {author} {\bibfnamefont {Charles~R.}\
  \bibnamefont {Doering}},\ }\bibfield  {title} {\enquote {\bibinfo {title}
  {Stochastic ratchets},}\ }\href {\doibase 10.1016/S0378-4371(98)00006-5}
  {\bibfield  {journal} {\bibinfo  {journal} {Physica A: Statistical Mechanics
  and its Applications}\ }\textbf {\bibinfo {volume} {254}},\ \bibinfo {pages}
  {1 -- 6} (\bibinfo {year} {1998})}\BibitemShut {NoStop}%
\bibitem [{\citenamefont {Luchsinger}(2000)}]{luchsinger00}%
  \BibitemOpen
  \bibfield  {author} {\bibinfo {author} {\bibfnamefont {Rolf~H}\ \bibnamefont
  {Luchsinger}},\ }\bibfield  {title} {\enquote {\bibinfo {title} {Transport in
  nonequilibrium systems with position-dependent mobility},}\ }\href@noop {}
  {\bibfield  {journal} {\bibinfo  {journal} {Physical Review E}\ }\textbf
  {\bibinfo {volume} {62}},\ \bibinfo {pages} {272} (\bibinfo {year}
  {2000})}\BibitemShut {NoStop}%
\bibitem [{\citenamefont {Lan{\c{c}}on}\ \emph {et~al.}(2001)\citenamefont
  {Lan{\c{c}}on}, \citenamefont {Batrouni}, \citenamefont {Lobry},\ and\
  \citenamefont {Ostrowsky}}]{lancon01}%
  \BibitemOpen
  \bibfield  {author} {\bibinfo {author} {\bibfnamefont {P}~\bibnamefont
  {Lan{\c{c}}on}}, \bibinfo {author} {\bibfnamefont {G}~\bibnamefont
  {Batrouni}}, \bibinfo {author} {\bibfnamefont {L}~\bibnamefont {Lobry}}, \
  and\ \bibinfo {author} {\bibfnamefont {N}~\bibnamefont {Ostrowsky}},\
  }\bibfield  {title} {\enquote {\bibinfo {title} {Drift without flux: Brownian
  walker with a space-dependent diffusion coefficient},}\ }\href@noop {}
  {\bibfield  {journal} {\bibinfo  {journal} {EPL (Europhysics Letters)}\
  }\textbf {\bibinfo {volume} {54}},\ \bibinfo {pages} {28} (\bibinfo {year}
  {2001})}\BibitemShut {NoStop}%
\bibitem [{\citenamefont {Krishnan}\ \emph {et~al.}(1992)\citenamefont
  {Krishnan}, \citenamefont {Singh},\ and\ \citenamefont
  {Robinson}}]{krishnan92}%
  \BibitemOpen
  \bibfield  {author} {\bibinfo {author} {\bibfnamefont {R}~\bibnamefont
  {Krishnan}}, \bibinfo {author} {\bibfnamefont {Surjit}\ \bibnamefont
  {Singh}}, \ and\ \bibinfo {author} {\bibfnamefont {GW}~\bibnamefont
  {Robinson}},\ }\bibfield  {title} {\enquote {\bibinfo {title}
  {Space-dependent friction in the theory of activated rate processes},}\
  }\href@noop {} {\bibfield  {journal} {\bibinfo  {journal} {Physical Review
  A}\ }\textbf {\bibinfo {volume} {45}},\ \bibinfo {pages} {5408} (\bibinfo
  {year} {1992})}\BibitemShut {NoStop}%
\bibitem [{\citenamefont {Van~Kampen}(1988)}]{vankampen1988}%
  \BibitemOpen
  \bibfield  {author} {\bibinfo {author} {\bibfnamefont {NG}~\bibnamefont
  {Van~Kampen}},\ }\bibfield  {title} {\enquote {\bibinfo {title} {Relative
  stability in nonuniform temperature},}\ }\href@noop {} {\bibfield  {journal}
  {\bibinfo  {journal} {IBM Journal of Research and Development}\ }\textbf
  {\bibinfo {volume} {32}},\ \bibinfo {pages} {107--111} (\bibinfo {year}
  {1988})}\BibitemShut {NoStop}%
\bibitem [{\citenamefont {Landauer}(1988)}]{landauer88}%
  \BibitemOpen
  \bibfield  {author} {\bibinfo {author} {\bibfnamefont {Rolf}\ \bibnamefont
  {Landauer}},\ }\bibfield  {title} {\enquote {\bibinfo {title} {Motion out of
  noisy states},}\ }\href {\doibase 10.1007/BF01011555} {\bibfield  {journal}
  {\bibinfo  {journal} {Journal of Statistical Physics}\ }\textbf {\bibinfo
  {volume} {53}},\ \bibinfo {pages} {233--248} (\bibinfo {year}
  {1988})}\BibitemShut {NoStop}%
\bibitem [{\citenamefont {Risken}(1984)}]{risken84:book}%
  \BibitemOpen
  \bibfield  {author} {\bibinfo {author} {\bibfnamefont {H.}~\bibnamefont
  {Risken}},\ }\href {http://books.google.de/books?id=7KXnQwAACAAJ} {\emph
  {\bibinfo {title} {The Fokker-Planck Equation: Methods of Solution and
  Applications}}}\ (\bibinfo  {publisher} {World Publishing Corporation},\
  \bibinfo {year} {1984})\BibitemShut {NoStop}%
\end{thebibliography}


%\bibliographystyle{plain}
%\bibliographystyle{elsarticle-num}

\end{document}